\documentclass[12pt]{iopart}
\usepackage{graphicx}
\usepackage{xspace}
\usepackage{amssymb}
\usepackage{soul}
\usepackage{tikz}
\usepackage{xcolor}

\renewcommand{\etal}{\emph{et al}.\xspace}

\usepackage{hyperref}
\usepackage{url}
\definecolor{lightred}{rgb}{1,0.85,0.88}
\hypersetup{colorlinks,breaklinks,
            linkcolor=blue,urlcolor=blue,
            anchorcolor=blue,citecolor=blue}

\begin{document}

\topical[Compound screening of soft matter by molecular simulations]
{Computational compound screening of biomolecules and soft materials
by molecular simulations}

\author{Tristan Bereau}

\address{Van 't Hoff Institute for Molecular Sciences and Informatics Institute,
University of Amsterdam, Amsterdam 1098 XH, The Netherlands}

\address{Max Planck Institute for Polymer Research, 55128 Mainz, Germany}

\ead{t.bereau@uva.nl}

\begin{abstract}
	Decades of hardware, methodological, and algorithmic development
	have propelled molecular dynamics (MD) simulations to the
	forefront of materials-modeling techniques, bridging the gap
	between electronic-structure theory and continuum methods. The
	physics-based approach makes MD appropriate to study emergent
	phenomena, but simultaneously incurs significant computational
	investment. This topical review explores the use of MD outside the
	scope of individual systems, but rather considering many
	compounds. Such an \emph{in silico} screening approach makes MD
	amenable to establishing coveted structure--property
	relationships. We specifically focus on biomolecules and soft
	materials, characterized by the significant role of entropic
	contributions and heterogeneous systems and scales. An account of
	the state of the art for the implementation of an MD-based
	screening paradigm is described, including automated force-field
	parametrization, system preparation, and efficient sampling across
	both conformation and composition. Emphasis is placed on
	machine-learning methods to enable MD-based screening. The
	resulting framework enables the generation of compound--property
	databases and the use of advanced statistical modeling to gather
	insight. The review further summarizes a number of relevant
	applications.
\end{abstract}
\submitto{\MSMSE}
\maketitle

\section{Introduction}
\label{sec:intro}

Ceder and Persson's \emph{Scientific American} article \emph{The Stuff
of Dreams} refers to the ``golden age of materials design,'' a new era
where computational methods---a mix of hardware and software
implementation of physical laws and equations---assist scientists in
designing new functional materials \cite{Ceder2013}. Designing better
materials means selecting a chemical composition that yields superior
materials properties. Traditional avenues have followed an Edisonian,
trial-and-error approach, by experimentally screening as many
compounds as possible---an approach that is typically both
time-consuming and costly, due in no small part to synthesis,
processing, and characterization. Computation offers a parallel route
to search for compounds with desired characteristics, where the
numerical solution of fundamental equations (e.g., the Schr\"odinger
equation) can make predictions before going to the laboratory. The
effort has gained momentum thanks to the development of computational
hardware, software, and database tools, demonstrating exceptional
potential to accelerate materials discovery in various fields
\cite{jain2013commentary, curtarolo2013high, pyzer2015high,
jain2016computational, ramprasad2017machine, Tkatchenko2020}.

There are good reasons to expand compound screening beyond the
experimental realm. While high-throughput screening can probe
impressive numbers of candidates, the requirements to synthesize,
process, and/or characterize large libraries of compounds typically
restricts the approach to particular systems and properties
\cite{Mishra2008, Mayr2009, Macarron2011, Potyrailo2011, Muster2011,
du2016microfluidics}. The computational route certainly also holds its
share of system and property limitations, but are alleviated by the
variety in resolutions, methods, and algorithms. Limitations may also
arise from the set of compounds accessible: synthesized drugs form a
minuscule subset of the chemical space of small organic molecules
\cite{Dobson2004}. While not all compounds are expected to be
necessary to satisfyingly interpolate the space, the level of
subsampling unfortunately leads to a lack of uniformity: a database
bias \cite{Hert2009, Lin2018}. Screening on the computer, on the other
hand, needs no synthesis---though its virtual analog, model
parametrization, often remains a challenge. More flexibility in
choosing compounds enables avenues to exhaustively enumerate small
subsets \cite{frederix2011virtual}, find efficient ways to build up
combinatorics \cite{Fink2007}, and select compounds using more
sophisticated strategies, for instance active learning
\cite{Warmuth2003}.

To remain robust across chemical space, computational methods must
rely on fundamental, broadly applicable physical laws and equations.
These physics-based methods---including the Schr\"odinger and
Kohn--Sham equations at the electronic-structure level and Newton's
classical equations of motion at the classical level---can make
predictions that are grounded in the corresponding physics. Even
classical simulations typically give rise to significant computational
costs, which had until recently limited their penetration into the
field of compound screening. Turning to density functional theory
(DFT), the recent development yet rapid adoption of high-throughput
schemes for various materials applications testifies to the escalating
role of computation in materials screening and discovery
\cite{curtarolo2013high, Saal2013, Jain2016, Himanen2019}.

While some fields have already benefitted strongly from computational
screening, others lag behind---such is the case for soft condensed
matter. Marked by weak characteristic interaction energies on par with
thermal energy, $k_{\rm B}T$, soft-matter systems embody a large class
of materials, including not only polymers, liquid crystals,
surfactants, colloids, but also biomolecules. When coupled to thermal
fluctuations, soft matter display fascinating phenomena, such as
spontaneous self assembly and mesoscopic architectures, simply
navigating a rich free-energy landscape \cite{doi2013soft}.
Fluctuations de facto require a careful consideration of entropic
effects, and adequate computational methods to sample the accessible
conformational space. Furthermore, soft-matter systems also typically
display poor scale separation, challenging multiscale-modeling
approaches \cite{peter2010multiscale}.

The challenges of modeling biomolecules and soft matter have largely
kept the field in a ``craftsmanship era.'' Scientific studies focus on
one or a handful of compounds, due to difficulties in parametrizing,
preparing, sampling, and analyzing the system. These aspects all stand
orthogonal to a screening strategy---automation reigns over the
high-throughput paradigm. It is thus no surprise that machine learning
and other data-driven techniques are rapidly penetrating the field of
soft materials \cite{ferguson2017machine, bereau2018data,
Jackson2019}. The rapid rise of high-throughput molecular simulations
is the topic of this review.

\subsection{Scope}
Compound screening is a vast, quickly evolving area that connects to
physics and chemistry, materials science, and even branches out to a
plethora of applications, from organic photovoltaics to
electrocatalysis to drug discovery to biomaterials \cite{Greeley2006,
Simon2010, Kitchen2004, Hachmann2011, Potyrailo2011}. Despite its
focus on biomolecular systems and soft matter, this compound-screening
review will exclude studies originating from experimental
data---arguably its largest subset. A large body of work has been
devoted to the utilization of experimental compound databases, notably
from quantitative structure--activity relationship (QSAR) methods in
drug discovery \cite{Hellberg1987, topliss2012quantitative,
le2012quantitative}. Instead this review will focus not only on
computational (\emph{in silico}) screening, but those generated from
\emph{physics-based} methods. Physics-based methods consist of a
hierarchy of multiscale-modeling methods, from quantum chemistry, to
empirical force-field-based molecular dynamics (MD), to particle-based
coarse-grained (CG) simulations, to continuum modeling
\cite{peter2010multiscale, tadmor2011modeling, vanderGiessen2020}.
They prevail in some key aspects essential to biomolecular materials
and soft matter, specifically the modeling of emergent phenomena and
entropy. Further, this hierarchy offers a conceptual bridge to the
funnel-like nature of compound screening: quickly screen with fast
methods and refine with more accurate models. 

Current computational limitations strongly limit a purely
quantum-chemical approach to a limited range of problems: primarily
isolated molecules or relatively small and homogeneous environments
\cite{szabo2012modern}. Classical MD simulations prevail for
biomolecules and soft matter, because of their ability to efficiently
sample the vast conformational space. For a history and overview of MD
simulations, we refer the reader to excellent books and reviews
\cite{binder1995monte, frenkel2001understanding, Karplus2002,
Hansson2002, rapaport2004art}. Though MD-based screening studies are
dominated by an atomistic resolution, CG models take an increasingly
large role, thanks to their more favorable computational load and
ongoing improvements in linking the lower resolution to the underlying
chemistry. This review will mostly revolve around \emph{spatial} CG:
particle-based models made of interaction sites (also called
superparticles or beads), which correspond to groups of atoms. On the
other hand, we will not touch upon methods that coarse-grain in
\emph{time}, due to (so far) limited impact on compound screening
\cite{Mori1965, Zwanzig1973, Tuckerman1992, Gear2003, Chodera2014}.

\subsection{Inverse problems in soft matter}
\label{sec:intro:inverse}
\begin{figure}[htbp]
    \centering
    \includegraphics[width=0.95\linewidth]{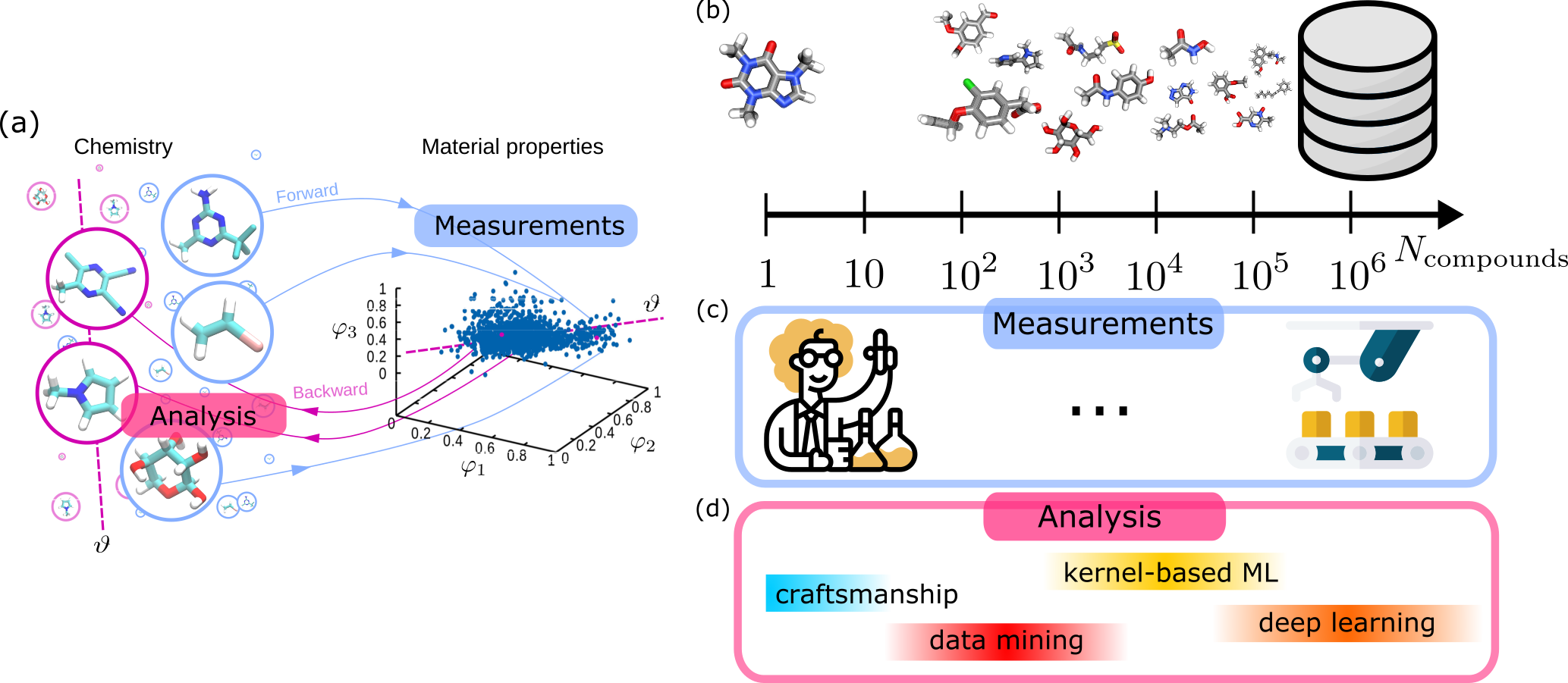}
    \caption{(a) Structure--property relationships are based on
    forward measurements and subsequent backward inference; (b)
    Analogous to length- and time-scales in materials modeling, the
    number of compounds---the data-scale---is an essential variable of
    compound-screening problems; (c) Measurements can only be
    performed manually for the lowest values of $N_{\rm compounds}$,
    but otherwise require automation. (d) Different scales of $N_{\rm
    compounds}$ are amenable to different types of statistical
    modeling. Part of the figure is adapted from
    \cite{bereau2016research}, under a Creative Commons Attribution
    (CC BY) license.}
    \label{fig:data_scale}
\end{figure}

A material, entirely determined by its chemical composition---but also
often its processing---will yield specific properties. Making
measurements, either by experimental techniques or
analytical/numerical calculations, boils down to establishing a
mapping between the material composition and its properties. This is
commonly denoted the \emph{forward} problem, and is illustrated in
Fig.~\ref{fig:data_scale}a \cite{bereau2016research}. Materials
design, on the other hand, aims at establishing the backward---or
inverse---mapping: identifying the adequate structure given properties
of interest. While the forward route is straightforward, there is no
experiment or equations of motion to directly probe the backward
problem. It instead typically requires solving an inverse problem:
from a (small) number of forward measurements, infer the function that
links chemistry to materials property. The notorious difficulty to
solve inverse problems also applies in materials discovery, and leads
to strenuous requirements on the number of measurements compared to
the size of the interpolation space \cite{kaipio2006statistical}.

Though commonly referred to as \emph{structure--property
relationships}, this terminology hides that the structure itself is
entirely determined by the material's chemical constituents. The
review by Sherman \etal clearly differentiates four different stages
in the design of soft materials: ($i$) chemical synthesis or
preparation leads to ($ii$) building blocks with effective,
coarse-grained interactions, which drive their assembly into ($iii$)
structures or morphologies, and imprint ($iv$) properties on the
macroscopic scale \cite{sherman2020inverse}. This
chemistry--building-block--structure--property framework does justice
to the complexity, heterogeneity, and large scale separation that
characterizes soft matter. 

The chemistry to building-block step, ($i \to ii$), is essential to
reduce the overwhelming vastness of chemical space \cite{Dobson2004,
Reymond2015} into a low-dimensional set of effective components with
coarse-grained interactions. This requires a thorough understanding of
the dominant driving forces: supramolecular interactions such as van
der Waals, electrostatics, or hydrogen bonds \cite{stone2013theory}.
Modeling has greatly taken advantage of building blocks by means of
top-down coarse-graining, which parametrize simple models based on key
phenomenological interactions, while staying close to the chemistry
\cite{noid2013perspective, Inglfsson2013}. The building-block to
structure step, ($ii \to iii$), has likely received the most
attention. Relevant work largely consists of improving our
understanding or finding practical routes at linking coarse-grained
interactions to self assembly. Notable examples include the directed
self assembly of diblock copolymer thin films using self-consistent
field theory \cite{Hannon2013}; The ``materials design engine,'' using
statistical mechanics as an automatic optimizer, with applications
including the folding of a polymer and the directed self assembly of
block copolymers \cite{Miskin2015}; Design principles for colloidal
self assembly with short-range interactions, establishing tight
restrictions on the relative strength of the favorable and unfavorable
interactions, as well as the number of components and energies
\cite{Hormoz2011}; A ``digital alchemy'' framework to control self
assembly by optimizing building blocks for a given target bulk
structure \cite{vanAnders2015}. The structure to property step, ($iii
\to iv$), has largely involved finite-element methods to optimize
material microstructures for specific design specifications, such as
acoustic, elastic, and photovoltaic properties \cite{Jain2014}.

At equilibrium an additional consideration may prove useful in
approaching inverse problems: the free-energy landscape. Central to
any soft-matter system, the free-energy landscape shapes the
self-assembly route, navigating down between conformational basins
toward a (local) minimum. The free-energy landscape also conditions
all observables, by its statistical weights over the conformational
space. In the context of solving the inverse problem, the free-energy
landscape thus stands as a powerful, physically meaningful
intermediary between chemistry and building-block constituents on the
one hand and structure/morphology and macroscopic properties on the
other.

How does changing the chemistry affect the free-energy landscape?
Various studies are tackling this question. Meng \etal reported the
free-energy landscape of clusters of attractive hard spheres,
including a detailed characterization of the rotational entropy
\cite{Meng2010}. Scaling up, the field of protein folding has led to
great insight into how the shape of the free-energy landscape impacts
a protein's properties---the famous funnel-like shape is
characteristic of many efficient folders \cite{Bryngelson1995,
Shakhnovich2006, Dill2012}. These developments further enabled the
design of new proteins, whose sequence and structure differ
significantly from naturally occurring proteins \cite{Kuhlman2004}.
Unfortunately not all free-energy landscapes display straightforward
shapes; self assembly often results from a competition between
conformational basins. Jankowski and Glotzer carefully studied the
assembly pathway of patchy particles to grasp the diversity of
possible final structures \cite{Jankowski2012}. 

Coarse-graining likely has a strong role to play in the context of
screening. As described below in Section \ref{sec:app:drugmem}, a
high-throughput study of drug--membrane thermodynamics linked
coarse-grained features of small molecules with their potential of
mean force of insertion in a lipid membrane \cite{menichetti2018drug}.
The results suggest that exploring the diversity of top-down CG
building blocks (step $ii$) fittingly \emph{simplified} the
structure--property relationship, making it easier to identify. CG
models evidently coarsen the underlying free-energy landscape, and
what could be criticized as a loss in accuracy or resolution can also
be seen as a decisive advantage to tackle the inverse problem.

The system-size limitations associated with MD simulations naturally
hinder the prospects of scaling up to genuine macroscopic properties.
The systems remain instead micro- to mesoscopic and focus on basic
structural, thermodynamic, and sometimes dynamical aspects. Their
particle-based nature also naturally lend themselves to starting from
the $(i)$ chemistry or $(ii)$ building-block steps.

\subsection{Data-scales}
\label{sec:data-scales}
One landmark property of most---if not all---materials is the large
dynamic range of relevant length- and time-scales. Microscopic
interactions lead to mesoscale architectures and morphologies, but
also conformational transitions and aging behavior. It is not
uncommon to observe phenomena spanning 10 or more orders of magnitude
for either scale: from sub-nanometer to meter, and from femtosecond to
seconds or more. Interestingly, these scales are relevant not only to
understand the intrinsic properties of the system, but also to
\emph{probe} it: both experimental techniques and computational
methods typically specialize in probing a (possibly small) subset of
these scales \cite{peter2010multiscale, tadmor2011modeling,
serdyuk2017methods}. For instance, quantum-chemical methods reign at
small length- and time-scales, but fall short much beyond the
nanometer- and picosecond-marks.

In this review we apply a similar conceptual framework to the
\emph{number of screened compounds}, $N_{\rm compounds}$. This
data-scale, unlike its other two counterparts, is not an intrinsic
variable---it is merely a practical consideration to help guide both
the forward-measurement and backward-inference processes. We refer the
reader to Figure~\ref{fig:data_scale} for an illustration:
establishing structure--property relationships (panel a) hinges upon
the number of compounds screened (panel b). As will be described in
Section~\ref{sec:app}, MD studies typically work in the range $1 \leq
N_{\rm compounds} \lesssim 10^6$, though steady progress will likely
rapidly push the upper bound. Working in higher regimes of the data
scale will on the one hand strongly impact \emph{requirements} on the
forward-measurement protocol (Figure~\ref{fig:data_scale}c), but on
the other hand \emph{permit} more sophisticated statistical-analysis
techniques (Figure~\ref{fig:data_scale}d). The data scale thereby
forms an essential pillar to guide a compound-screening study, both to
generate a database and garner insight from it.

\section{Computational high-throughput paradigm}
\label{sec:hts}
\begin{figure}[htbp]
	\centering
	\includegraphics[width=0.8\linewidth]{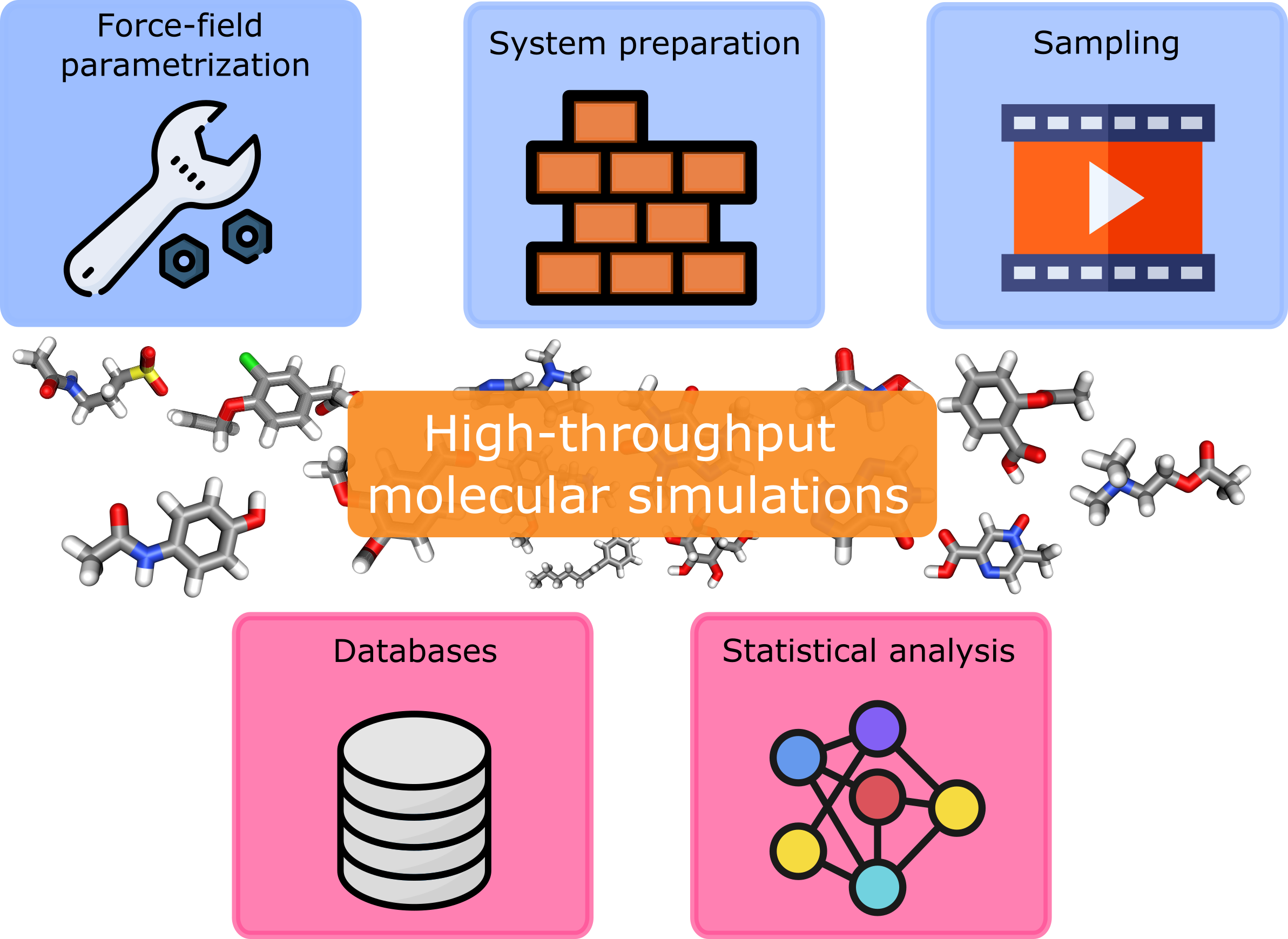}
	\caption{Protocol for high-throughput molecular simulations.
		Requirements include automated force-field parametrization
		schemes, system preparation, and efficient sampling (top;
		blue). It enables the generation of compound databases and
		statistical analysis to gather insight (bottom; pink).}
	\label{fig:hts_protocol}
\end{figure}

Before moving onto applications (Section \ref{sec:app}), we first
describe the forward-measurement requirements and backward-analysis
possibilities that a computational high-throughput paradigm both
impose and enable, sketched in Figure~\ref{fig:hts_protocol}. The
forward-measurement steps necessary to build the compound
database---the blue boxes in Figure \ref{fig:hts_protocol}---embody
the computational analog of a laboratory's high-throughput screening
experiment. The framework demands a strict and homogeneous protocol
across compounds for two reasons: ($i$) it yields a \emph{consistent}
database amenable to extracting structure--property relationships; and
($ii$) it is practically convenient for automation purposes.  The
present section describes the various aspects of running MD
simulations under these constraints.

When possible, the examples will be borrowed from the biomolecular and
soft-matter fields. In other cases however, examples from other
fields---in particular chemistry and hard condensed matter---may prove
insightful of where developments may be headed.

\subsection{Force-field parametrization}
The scope and level of refinement of a number of biomolecular force
fields attest to the remarkable developments in the
molecular-simulation field: some of them are decades in the making,
amounting to thousands of finely tuned parameters, and have endured
relentless evaluations \cite{maple1988derivation,
halgren2001polarizable, wang2001biomolecular, ponder2003force,
mackerell2004empirical, wang2017building}.  Unlike more empirical
methods (e.g., statistical scoring in drug discovery), the
physics-based nature of force fields grounds the model in the physics
considered. It relies on specific potentials that encode relevant
interactions \cite{halgren1992representation, tkatchenko2012accurate,
van2016beyond}. Unfortunately force fields are difficult beasts to
tame: their complexity can easily make any (re)parametrization for new
compounds laborious, because they do not always offer systematic
strategies.

Automated force-field parametrization is an old idea that is difficult
to practically implement. Why is that? Quantum mechanics ought to
provide us with a sure-fire way to derive classical potentials.
Unfortunately the physics encoded in force fields is rather limited:
for instance, most force fields are not explicitly polarizable. The
limited physics of the model clouds the relationship to quantum
mechanics and instead warrants a parametrization based on experimental
properties. Major biomolecular force fields, such as CHARMM and OPLS,
typically use a combination of reference information to parametrize
across the chemical compound space (CCS; more on that in
Section~\ref{sec:sampling:comp}) of drug-like small molecules: like
others the CHARMM general force field (CGenFF) uses quantum mechanics
to optimize charges and bonded interactions, while Lennard-Jones
parameters rely on experimentally determined liquid density and heat
of vaporization \cite{vanommeslaeghe2015charmm}. The need for
experimental quantities can be problematic, and is alleviated by
identifying chemical groups or fragments found in previously analyzed
molecules. The gradual incorporation of model compounds allows CGenFF
to broadly interpolate across a large subset of CCS, while retaining
high fidelity of structural and thermodynamic properties. A similar
strategy has been applied by OPLS \cite{Harder2015, Roos2019}, GROMOS
\cite{Malde2011}, and AMBER \cite{Wang2006automatic}.

Arguably the incorporation of experimental data in a
computational-screening pipeline is unfortunate: experimental data are
limited to a minuscule subset of CCS, and it might well defeat the
purpose of a virtual compound-discovery study. Despite their broad
coverage of CCS, the above-mentioned biomolecular force fields largely
avoid this issue by sharing and reusing information between molecules.
The piece of information that is typically shared is the \emph{atom
type}. Beyond the chemical element itself, it represents the atom in a
molecule given a local environment, for instance an sp$^2$ carbon in
an alkene. The more chemically specific, the better---in other words
the larger incorporation of neighboring atoms will more precisely
characterize the local environment, and offer all the more resolution.
The above-mentioned automated force-field strategies primarily aim at
selecting the right atom types, and extract the corresponding
parameters from a database. While these atom types have historically
been handcrafted by chemical intuition, ongoing efforts aim at
generalizing its concept using more robust annotators. For instance,
the Open Force Field Initiative is applying so-called direct chemical
perception by the use of SMIRKS patterns---linear notations encoding
atoms and bonds \cite{Mobley2018}.

The tendency to encode increasingly many atom types begs the question:
is there a continuum limit? In effect this is precisely what is probed
by machine learning (ML) models that span (subsets of) CCS. While we
defer a broader discussion on the topic to Section \ref{sec:data}, we
note that kernel-based methods, such as Gaussian process regression
(GPR), assume and enforce smoothness of the input space by the kernel
function \cite{rasmussen2004gaussian}. It leads to a continuum
description of a so-called \emph{atom-in-molecule} representation, a
concept strongly utilized in hard condensed matter
\cite{tkatchenko2012accurate}. ML models learn a smooth interpolation
between many-body atom-in-molecule representations and a target
property of interest. ML has rapidly demonstrated impressive
capabilities to interpolate increasingly large subsets of the CCS to
complex electronic properties. Examples include atomization energies
\cite{rupp2012fast}, dipole polarizability tensor \cite{Wilkins2019},
and multipole electrostatic coefficients
\cite{bereau2015transferable}.

How do we incorporate ML models into force fields? One straightforward
approach is to work simultaneously with both: physics-based force
fields encode the functional forms and asymptotes that we know, while
ML models predict composition- and conformation-specific environments.
This approach can lead to excellent accuracy and transferability,
reproducing highly accurate coupled-cluster calculations across
several molecular datasets, and without the need for any
reparametrization \cite{bereau2018non}. Li \etal have used ML models
to predict quantum-mechanical properties, used as input for a
polarizable force field, and match liquid-state observables
\cite{li2017machine}. In both cases the high-resolution of the
physics-based models---they are both explicitly polarizable---enable a
purely \emph{ab initio} parametrization. 

The more ML-centric alternative is to let go of functional forms
entirely. Several applications show that this can lead to excellent
many-body ML potentials for a variety of molecules and materials
\cite{bartok2010gaussian, behler2016perspective, chmiela2017machine}.
Moving beyond single systems and toward subsets of CCS is still a
subject of ongoing research: most of these approaches have so far
focused on a careful interpolation of the conformational space, and
the compounded interpolation of composition requires significant
adaptations (Section~\ref{sec:sampling:comp}). We point out the ML
neural network potential ANI as a notable example in this direction
\cite{Smith2017}. We also note the challenge of accurately modeling
long-range interactions, for instance by appropriate physically
inspired kernels \cite{grisafi2019incorporating}.

Going down in resolution, developing CG models takes the simulator
down either one of two main tracks: top-down or bottom-up
\cite{noid2013perspective}. The top-down approach, which builds from
phenomenological considerations, may turn out easier to automate in
the case that there is a straightforward link between the reference
information and the interaction potential. A variety of powerful
models have been developed in the past, and we turn the interested
reader to relevant reviews \cite{noid2013perspective, Inglfsson2013}.
Consider the popular CG Martini force field for biomolecular systems
\cite{marrink2013perspective}. The automated CG Martini
parametrization scheme can read in any small organic molecule,
optimize a mapping using a set of heuristics, and predict a chemical
fragment water/octanol partitioning coefficient from a neural network
for each bead type \cite{bereau2015automated}. Bead types of CG models
can be further redefined to best accommodate for the diversity of
compounds in the CCS \cite{Kanekal2019}. On the other hand, the
bottom-up route starts from microscopic information of a
higher-resolution simulation. Systematic parametrization schemes
exist, such as iterative Boltzmann inversion or force matching,
accompanied by convenient software platforms \cite{Rhle2009,
Dunn2017}. Aside from the CG potentials, bottom-up strategies can
strongly benefit from a more systematic optimization of the mapping
itself \cite{Chakraborty2020, Foley2020}. Combinations of
structure-based CG and ML have recently sparked interest and are
quickly enabling new avenues, see below Section~\ref{sec:data:kernel}.

\subsection{System preparation}
\label{sec:sysprep}
System preparation for an MD study has two main tenets: ($i$) the
initial configurations and ($ii$) the procedure to run the simulation
and compute observables (e.g., structural parameter or free energy).
Controlling the latter is typically relatively easy, as it often boils
down to applying the same simulation pipeline. Building initial
configurations in an automated and consistent way, on the other hand,
can require more sophisticated approaches: A screening study that
focuses on protein--ligand binding must first dock every single
compound in the protein pocket. Beyond the proper geometric alignment
of the ligand, the condensed phase of a liquid calls for packing of
the molecules involved, and thus a delicate placement to avoid steric
clashes.  This has led to a variety of tools to initialize
condensed-phase, soft-matter systems: Mart\'inez \etal designed
PACKMOL to create simple liquids, mixtures, and more complex
architectures, such as micelles and lipid bilayers \cite{Martnez2009};
Polymer Modeler is a polymer chain builder \cite{polymermodeler};
CHARMM-GUI is a sophisticated web server to facilitate the initial
configuration of biomolecular systems, such as solvated proteins, and
phospholipid membranes \cite{Jo2008}; the INSANE script sets up
complex phospholipid-membrane mixtures for the CG Martini force field
\cite{wassenaar2015computational}; MemProtMD elegantly prepares CG
configurations of membrane proteins by \emph{self-assembling} the
phospholipid membrane around the experimentally resolved protein
structure (Section \ref{sec:app:membprot})
\cite{newport2019memprotmd}; both the Python-based MoSDeF and Hoobas
frameworks offer extensible molecular-building capabilities (e.g.,
patchy DNA-grafted colloids in Hoobas), and the use of Python allows
for deeper integration of system initialization and
simulation/analysis \cite{girard2019hoobas, Summers2020}.

\subsection{Sampling}
\label{sec:sampling}
Sampling lies at the heart of molecular simulations: both molecular
dynamics (with appropriate thermostat) and Monte Carlo simulations
implement efficient importance-sampling algorithms to navigate a
representative subset of the conformational space
\cite{frenkel2001understanding}.  But sampling takes on a whole new
dimension in the context of this review: not only does a simulation
aim at sampling conformational space, compound screening is
\emph{also} a sampling problem---this one in compositional space. Here
we limit our overview to recent methods that aim at sampling either
space. The use of similar techniques to tackle both spaces is no
coincidence, it highlights their resemblance and the associated
sampling challenges.

\subsubsection{Conformational sampling.}
\label{sec:sampling:conf}
The conformational space represents the structural distribution
function of the system. A collection of $N$ particles will give rise
to a continuous $3N$-dimensional space of microstates. The statistical
ensemble used to probe the system biases the weighting of the states,
e.g., the Boltzmann distribution in the canonical ensemble. This bias
means that not all microstates contribute equally, and instead an
efficient conformational-sampling strategy should focus only on the
more important ones.

More conformational sampling is almost always desired: simulating
larger and more complex systems potentially opens up new insight
unattainable before, but also helps testing for convergence
issues \cite{neale2011statistical, Shaw2010}. Limited computational
resources limit how long the simulations can be, and instead offset
many efforts in sampling more efficiently. Several excellent reviews
cover the vast and rich area of enhanced-sampling techniques
\cite{Abrams2013, Bernardi2015, Valsson2016, Camilloni2018}. 

ML, and in particular deep learning, has opened up a number of new
avenues in terms of facilitating conformational sampling
\cite{ferguson2017machine}. For instance, autoencoders display an
architecture that is prone to enhanced sampling: its symmetric bow-tie
network, while simply aiming at reconstructing the input sample,
forces an information bottleneck in the so-called \emph{latent space}.
Describing a system through this reduced dimensional latent space
bridges naturally to the use of collective variables in enhanced
sampling. A famous variant to autoencoders, the variational
autoencoder, uses a variational approach to learn the latent
representation, resulting in both a generative model and a smooth
latent space that enables interpolation \cite{kingma2013auto}. Various
studies have leveraged the architecture of a (variational) autoencoder
to learn a low-dimensional latent representation of the input
conformational space \cite{Chen2018, Sultan2018} or extract the
long-time kinetics \cite{Wehmeyer2018}. The added accuracy one can
gain by using ML often comes at the cost of interpretability: how do
we express the latent-space dimensions---the collective variables---in
terms of simple, physically meaningful coordinates? Ribeiro \etal
proposed to iteratively refine a set of proxy reaction coordinates
that best emulates the latent-space distribution \cite{Ribeiro2018}.

Other approaches do away with collective variables, and instead use
unsupervised learning as a way to chart a low-dimensional free-energy
surface. Chiavazzo \etal have devised a method that iteratively
proceeds between MD and nonlinear manifold learning techniques to
expand the system away from regions already explored
\cite{Chiavazzo2017}. Expanding conformational space using
dimensionality reduction was also proposed by Kukharenko \etal
\cite{kukharenko2016using}. They used the multidimensional-scaling
scheme sketch-map \cite{Ceriotti2011} to project the points and
initiate swarms of simulations from sparsely (but existing) sampled
regions. The generation of molecular configurations that have not been
previously sampled was subsequently proposed by means of a loss
function that combined an autoencoder reconstruction loss and the
sketch-map cost function \cite{Lemke2019}. The combination of the two
approaches effectively appears to achieve features in line with the
variational autoencoder: the data-driven learning of a smooth
latent-space distribution, coupled to a generative model.

Beyond techniques aiming at enhancing the conformational space
sampled, others have tried to blend in qualitative external
knowledge---a prior of sorts---to drive the molecular dynamics. Perez
\etal employed Bayesian inference to guide protein-folding MD from
coarse physical knowledge, such as ``form a hydrophobic core''
\cite{perez2015accelerating}. Folding times were reduced by several
orders of magnitude, illustrating that the body of insight about
protein folding can be leveraged to speed up protein simulations. This
example illustrates well the dichotomy between what is systematic
(e.g., algorithms) and what is not (e.g., our intuition), and the
Bayesian scheme provides a formalism to bridge the two approaches.
Strategies to blend numerical methods or algorithms with heuristic
prior knowledge is bound to be useful in other areas.

\subsubsection{Compositional sampling.}
\label{sec:sampling:comp}
The chemical compound space (CCS)---the space of all possible
molecules or compounds---differs from the conformational space in at
least two major ways: First, its discreteness. Conformational space
permits continuous transformations between any pair of microstates. On
the other hand, different molecules cannot be arbitrarily close,
because of basic chemical rules (e.g., valency). In other words, very
few spatial arrangements of atoms will lead to chemically stable
compounds. Although there are computational treatments to continuously
transform molecules (\emph{vide infra}), the common setting is to
dedicate different simulations for different molecules. 

The second defining feature of CCS is its size: the dimensionality of
the space is not a simple function of the number of particles. Natural
proteins can be built by combinations of 20 amino acids, meaning that
there are $20^{n}$ unique sequences of chain length $n$. For very
short peptides of length $n = 10$---barely long enough to stabilize
any secondary structure---this already leads us to a space of
$10^{13}$ compounds. The increased variety of chemical groups in
synthetic polymers will evidently yield a much larger CCS. Now
consider small-drug like molecules that obey Lipinski's ``rule of
five''---restricting the molecular weight, hydrophobicity, and number
of hydrogen bonds---which capture the physicochemical properties of
most orally active drugs \cite{Lipinski2004}, its space is estimated
at $10^{60}$ chemically stable molecules \cite{Dobson2004}. There are
not enough carbon atoms in the universe to synthesize all of them!
What can we do, then? Just like microstates, not all molecules are
made equal---most will yield uninteresting properties. Focusing on the
ones with desired properties is precisely the answer to solving the
inverse problem (Section \ref{sec:intro:inverse}).

While overwhelmingly large, important steps in better grasping the
size and scope of the CCS of drugs have been made. Reymond and
coworkers have sidestepped the minuscule, inconsistent collection of
synthesized drug-like molecules by instead constructing them
algorithmically \cite{Reymond2015, Fink2007}. Graph-based methods
combined with valency rules offer a systematic way to enumerate large
subsets of CCS---most of which have never been synthesized. The
so-called ``generated database'' (GDB) enumerates a dense coverage of
molecules containing a set of elements up to a threshold in number of
heavy atoms: the GDB-17 contains $10^{11}$ molecules up to 17 heavy
atoms of C, N, O, S, and halogens \cite{Ruddigkeit2012}. Beyond their
identity, computing \emph{properties} of these dense subsets has
subsequently been subject to much activity, because they enable the
training of ML models (Section \ref{sec:data}). The GDB has been used
for the calculation of electronic properties, typically from
density-functional theory (DFT), of increasingly many compounds: Rupp
\etal calculated the atomization energy of $7 \cdot 10^3$ molecules
\cite{rupp2012fast}; Ramakrishnan \etal computed various electronic
properties for $1 \cdot 10^5$ molecules
\cite{ramakrishnan2014quantum}; and Hoja \etal more recently reported
a database of $4 \cdot 10^6$ molecules \cite{hoja2020qm7}.

When tackling the exploration of CCS, coarse-graining can offer
significant advantages. Top-down, phenomenological CG models focus the
modeling on the essential ingredients or driving forces at play
\cite{noid2013perspective}. This minimalistic approach can lead to
generic---if not universal---behavior that broadly applies to many
systems. One famous example is the Kremer--Grest polymer model
\cite{Grest1986, Kremer1990}. Zhang \etal demonstrated that a melt of
this phenomenological model can broadly be backmapped to many
different types of homopolymers \cite{Zhang2015}. Everaers \etal
recently matched the generic large-scale behavior of Kremer--Grest
simulations to chemistry-specific experiments via the Kuhn length
\cite{Everaers2020}.

\begin{figure}[htbp]
	\centering
	\includegraphics[width=0.8\linewidth]{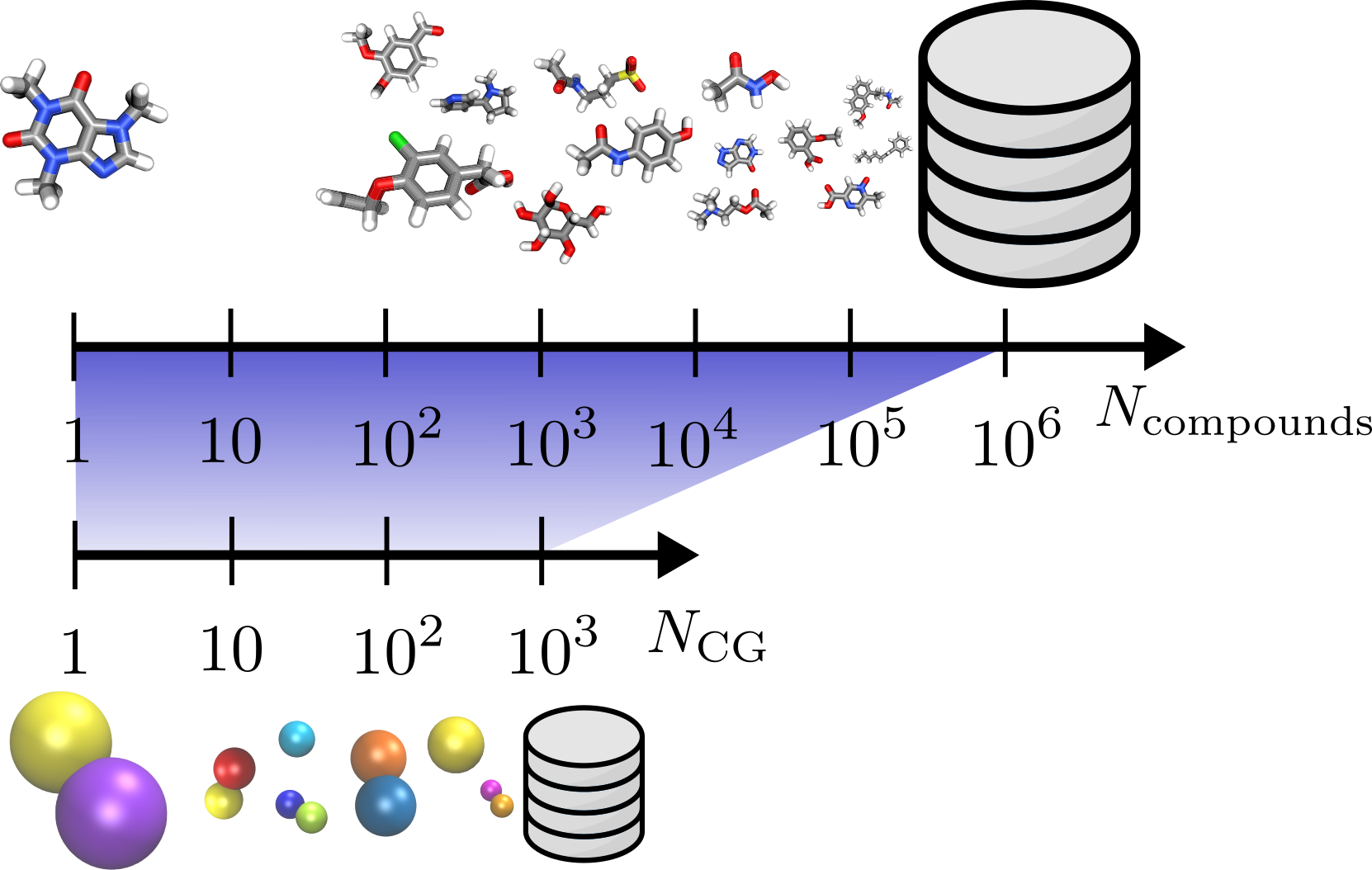}
	\caption{Transferable coarse-grained models can reduce the size of
		chemical compound space: fewer coarse-grained (CG) compounds are
		required to probe a subset of chemical space. They make use of a
		finite set of bead types to introduce a degeneracy in the CG
		representations of chemical compounds \cite{bereau2015automated}.}
	\label{fig:data_scale_cg}
\end{figure}

While the link between top-down CG models and the underlying CCS often
remains qualitative, there can be approaches to establish it. Many of
these top-down models are transferable, in that they define a limited
set of interactions of \emph{bead types} to encode the variety of
chemical groups. In case of the popular Martini model the bead types
roughly span the hydrophobicity scale \cite{marrink2013perspective}.
This limited chemical resolution means that molecules alike will often
map to the same CG mapping.  This critically introduces a degeneracy
in CG representation of small molecules, and effectively a
\emph{reduction} in the size of CCS. Figure \ref{fig:data_scale_cg}
illustrates the use of Martini for small molecules: it can lead to a
reduction in chemical space by roughly 3 orders of magnitude. The
mapping from molecules to CG representations is straightforward to
establish using automated parametrization schemes of GDB-type
libraries \cite{bereau2015automated, automartini}. This reduction of
the size of CCS can be applied to significantly boost the compound
screening of thermodynamic properties---one example will be covered in
the context of drug--membrane interactions, Section
\ref{sec:app:drugmem}. 

Beyond mere enumeration or serendipitous picks, there are more
efficient ways to explore CCS. Virshup \etal devised an algorithm to
stochastically grow an initial set of compounds to maximally diversify
it, restricted to specific properties (e.g., drug-likeness)
\cite{Virshup2013}. They reported a library of $10^4$ compounds
representative of the GDB-13, yielding a $10^4$ reduction factor while
retaining its diversity. Such an approach is likely to go hand in hand
with the training of ML models, which require a good balance between
chemical similarity and a representative coverage of the interpolation
space.  At the other end of the spectrum, Hoksza \etal presented the
{\sc Molpher} framework, which provides a (discrete) path in chemical
space between a pair of compounds \cite{Hoksza2014}. It performs a
series of simple structural molecular changes, such as atom addition
or removal, from start to target molecule.

Other approaches at sampling CCS emphasize the (bio)chemistry or
physics of navigating across molecules. Taking inspiration from nature
has led to the adaptation of Darwinian-type directed evolution
\cite{joyce1992directed}. Computational directed evolution has so far
mostly been applied to protein design, and more specifically to
enzymes \cite{Chowdhury2019}. Leveraging the aptness of computational
physics to perform importance sampling, a Markov Chain Monte Carlo
scheme can efficiently sample across CCS
\cite{hoffmann2019controlled}. Closer to reproducing a laboratory
experiment, Wang \etal implemented an \emph{ab initio} nanoreactor,
leading to spontaneous chemical reactions and the formation of
molecules through a variety of pathways \cite{Wang2014}. Such a
computational setting holds great promise in studying in more detail
the origins of life \cite{Meisner2019}.

While most of these approaches tackle CCS in its discrete form,
continuous explorations may well prove extremely strategic. However,
connecting compounds in a continuous manner requires some craft. One
notable example is the alchemical transformation, a powerful tool in
statistical mechanics to compute free-energy differences
\cite{Mobley2012}. It relies on a crucial property: state functions do
not depend on the path taken, and instead permit
non-physical---alchemical---interpolations between two compounds (more
on this in Sections \ref{sec:app:ligands} and
\ref{sec:app:solvation}).  A corresponding framework can be used to
compute \emph{ab initio} energy gradients and other changes in
properties upon local changes in CCS \cite{AnatolevonLilienfeld2009,
Balawender2013}. Aside from the relevant materials properties, the
inclusion of \emph{derivatives} may help in more efficiently mapping
structure--property relationships \cite{toBaben2016}. 

Another strategy to circumvent the discreteness of CCS consists of
imposing a continuous proxy. Such a proxy will enable
continuous-optimization schemes, thereby facilitating molecular
design. Wang \etal employed a linear combination of atomic potentials
to establish a continuous property landscape \cite{Wang2006}. In a
similar vein, von Lilienfeld \etal relied on an energy functional
based on the nuclear and electronic chemical potential
\cite{vonLilienfeld2005}. With the advent of deep learning, new
solutions have been proposed: G\'omez-Bombarelli \etal used a
variational autoencoder (covered in Section \ref{sec:sampling:conf})
to not only reduce the CCS, but more importantly to smoothen it
\cite{GmezBombarelli2018}. Built in the variational autoencoder, the
representation of the latent space allows a continuous exploration of
the CCS. The architecture was connected to a surrogate model, whose
objective was to predict a target property in the reduced latent
space, enabling continuous optimization. This active-learning,
Bayesian-optimization approach has lately been applied in the context
of soft-matter systems by Shmilovich \etal, as described in Section
\ref{sec:app:sapep} \cite{shmilovich2020discovery}.

\subsection{Data infrastructure}
Assuming all technical requirements permit MD simulations at high
throughput, the question arises: what to do with the data? Handling
large collections of MD simulations can easily require extensive
storage solutions. More importantly, it poses the problem of data
\emph{sharing}---not only between group members and collaborators, but
across the community at large. Recent cultural shifts in science are
increasingly encouraging the dissemination of research data. A
collaborative and open-source approach to scientific endeavors can
strongly accelerate the pace of research \cite{Woelfle2011}. Databases
of experimentally determined materials properties, for instance for
polymers, can prove invaluable to extract structure--property
relationships and assist in designing better materials \cite{Huan2016,
Audus2017, Barnett2020}.

What to do, then, to publish large collections of MD simulations? An
increasing number of online repositories dedicated to hosting
scientific data have come about, Zenodo \cite{zenodo_url}, figshare
\cite{figshare}, or the Open Science Framework \cite{osf}, to name but
a few. These databases are generic in that they are agnostic to the
type of scientific data, unlike, say, the Protein Data Bank (PDB),
which specializes in biomacromolecular structures \cite{pdb}. The next
question is the data format. One straightforward solution is to simply
compress all the input and output files of a set of MD trajectories
and upload them as is---a strategy our group adopted to publish
hundreds of umbrella-sampling MD trajectories
\cite{hoffmann2020molecular}. This lets anyone freely access the data,
but presents caveats. Notably, ($i$) it does not facilitate automated
strategies to search and collect information about the data, and
($ii$) the input/output formats are tied to MD software used to
generate the simulation trajectories. This is more formally denoted by
a lack of data labeling---or \emph{metadata}---and data normalization,
respectively. The convenient access, retrieval, and categorization of
heterogeneously generated data is key to assemble large databases,
amenable to training ML models (more on that in
Section~\ref{sec:data}). Such a framework has been formalized by the
FAIR principles: data that is Findable, Accessible, Interoperable, and
Reusable \cite{Wilkinson2016}. The new era of computational materials
design mentioned in the Introduction is in no small part made possible
by a robust data infrastructure in materials science
\cite{draxl2020big}. Publishing large FAIR datasets is becoming
increasingly widespread, thanks to solutions like the Materials Data
Facility \cite{Blaiszik2016}. The development of a number of
data-infrastructure platforms, such as NOMAD and the Materials
Project, strive to label electronic-structure calculations by detailed
metadata, parse many codes and normalize the input and output
information, and offer access via a webpage or a programmatic
interface \cite{Jain2013, draxl2018nomad}. Several consortia are
working their way toward more robust data infrastructures for
molecular simulations, including OpenKIM \cite{Tadmor2011,
Tadmor2013}, MOLSSI \cite{molssi}, and FAIR-DI \cite{fairdi}. Recent
examples show that the interconnection of specialized databases can
help automate the metadata annotation process, as will be described in
Section \ref{sec:app:membprot}.

\subsection{Data analysis}
\label{sec:data}
Once the difficult task of generating MD-based compound databases is
over, a second one starts: the data analysis. Here we will rely on the
concept of data-scale, already introduced in Section
\ref{sec:data-scales}. Figure \ref{fig:data_scale} illustrates that
the number of compounds largely determines the type of statistical
modeling. This constraint stems from the expressivity of a statistical
model, which depends largely on the number of parameters of the
architecture and dimensionality of the representation, which
themselves require larger training set sizes. We structure what
follows in terms of the data-scale by means of the variable
$N_\textup{compounds}$, from the traditional setting of craftsmanship,
to data mining in the low-data regime, to kernel-based ML methods, to
deep learning.

\subsubsection{Craftsmanship.}
Working in a regime $N_\textup{compounds} \sim 1$ leaves little room
for data-driven analysis methods. It instead embodies the traditional
setting of gathering insight driven by physical theories, experiments,
prior computer simulations, or simply intuition.

\subsubsection{Data mining.}
Moving up to $N_\textup{compounds} \gtrsim 10$ can offer enough
information to systematically search for simple structure--property
relationships. The low number of samples puts a strong limit on the
dimensionality of the sample information---the descriptors. Relating
low-dimensional descriptors to materials property has enjoyed great
attention for decades, embodied for instance by so-called quantitative
structure--property relationships (QSPR) \cite{le2012quantitative,
Lo2018}. QSPR is a well-established, powerful method to functionally
relate chemical structure to property. Applications include largely
drug discovery \cite{swift2013back, zhang2017machine}, but we also
note other soft-matter systems, such as the self assembly of
conjugated oligopeptides (more on that in Section \ref{sec:app:sapep})
\cite{Thurston2018} and the tribology of functionalized, lubricating
monolayer films \cite{Summers2020}. QSPR relies on a set of
descriptors, typically combined using a (multivariate) linear fit.
More recent applications have turned to using the kernel trick to
convert a non-linear problem into a linear one, support vector
machines can then highlight the most important descriptors, and we
further note the increasing use of artificial neural networks
\cite{swift2013back, zhang2017machine}. Practically however, these
data-mining models tend to be less limited by algorithmic developments
than by the data itself: small values of $N_\textup{compounds}$ can
easily lead to a large dependence to the training set. This aspect
calls for particular attention to \emph{model generalization}: how
similar do the predicted molecules need to be from the training set
\cite{swift2013back}. 

A more recent take on the functional discovery of structure--property
relationships brings us to learning more complex equations.
Compressed-sensing methods extend QSPR to expand the complexity of the
functional relationships tested. They rely on a large combinatorial
consideration of trial candidate equations, and a greedy $l^1$-norm
optimization scheme to minimize the number of non-zero coefficients.
Examples include the symbolic regression of nonlinear dynamical
systems \cite{Brunton2016} and equations from the \emph{Feynman
	Lectures on Physics} \cite{Udrescu2020}. Ghiringhelli \etal used least
absolute shrinkage and selection operator (LASSO) to extract
functional relationships between descriptors that can accurately
classify between zinc blende and rocksalt semiconductors
\cite{ghiringhelli2015big}. Ouyang \etal refined the approach using
the sure independence screening and sparsifying operator (SISSO),
which hierarchically searches for combinations of descriptors
\cite{ouyang2018sisso}. Rather than building a single surrogate model
aimed at explaining the entire dataset, another method called subgroup
discovery focuses on coherent homogeneous subsets. Goldsmith \etal
revisited the zinc-blende/rocksalt semiconductor problem and
identified separate regions with strict constraints
\cite{goldsmith2017uncovering}. These models are of particular
interest at a time where ML models are increasingly criticized for
their lack of interpretability: identifying the explicit role of the
input variable in the structure--property mapping.

By and large, these approaches aim at capturing the \emph{essential}
variables or descriptors that dictate the target property. This
dimensionality reduction aims at garnering insight into the problem at
hand, ideally by visualizing how the minimal set of descriptors link
to the property. The systematic construction of reduced dimensional
representations is a vast field, one that naturally connects to
unsupervised-learning techniques \cite{Ceriotti2019}. 

\subsubsection{Kernel-based supervised learning.}
\label{sec:data:kernel}
The regime $N_\textup{compounds} \gtrsim 10^3$ is amenable to the
optimization of much more expressive models. These are often called
surrogate models: they aim at learning the (oftentimes complex)
relationship between input and output parameters, so as to yield a
computationally efficient prediction. These models strive for accuracy
and generalization: to make a precise prediction over a large
interpolation domain. At best, the accuracy of the estimation can be
on par with the reference method \cite{faber2017prediction}. We refer
the reader to several excellent reviews on the use of (kernel-based)
ML for molecular systems \cite{pyzer2015high, behler2016perspective,
ramakrishnan2017machine, ferguson2017machine, ramprasad2017machine}.
Compared to QSPR methods, ML methods are free of fixed functional
forms, and instead offer flexible interpolation between training
points in a high-dimensional feature space
\cite{rasmussen2004gaussian,rasmussen2006gaussian}. ML models exploit
similarity in several ways: they first impose a \emph{metric},
allowing us to measure distances in CCS, a critical ingredient to both
explore and sample from that space (Section~\ref{sec:sampling}).
Similarity is explicitly assumed by enforcing smoothness between input
space and target property---an aspect that helps interpolate between
training points. 

Naturally, ML is not free of pitfalls.  The application of ML to
materials modeling---and more specifically to molecular
systems---requires domain knowledge. To be competitive, an ML model
should outperform an ambitious baseline: our own understanding of
physics and chemistry! An appealing strategy is to \emph{construct}
physics or chemistry inside ML models---an aspect we outline below.

The increased expressivity of ML relies on the use of
higher-dimensional input information, \emph{representations}, rather
than mere descriptors. Representations offer a more
detailed---many-body---description of the system, such as a molecule
or an atom in its local environment \cite{rupp2012fast,
bartok2017machine, faber2018alchemical, von2018quantum}. A
higher-dimensional representation also means more difficulties in
probing how broadly the ML model can be deployed: at which point does
it start extrapolating? How will we know? While there are many facets
to these questions, one crucial piece of information we can take
advantage of is the \emph{underlying physics}. Given that my system
obeys a conservation law or symmetry, can we constrain an ML model to
satisfy this constraint \emph{a priori}? The need to account for
physical symmetries was recognized early on \cite{Behler2007}. The
Noether theorem states that symmetries in a physical system lead to
conservation laws and invariants. Empirically learning these
invariants often requires significant amount of training
data---encoding them in the representation or the ML architecture can
lead to significant learning improvement
\cite{huang2016communication}. As a result, translation, rotation, or
(when applicable) permutation invariance often form the basic
requirements for ML representations. Symmetries can be added to the
kernel \emph{itself}, notable examples include the learning of vectors
by covariant kernels \cite{glielmo2017accurate} or energy-conserving
force fields via the Hessian \cite{chmiela2017machine,
bartok_gaussian_2015, Scherer2020}. Additional constraints can be
added as well, for instance a decomposition ansatz when the target
property lumps several terms, useful to decompose reference forces
\cite{bartok2010gaussian}, atomic dipole moments \cite{Veit2020}, or
free energies \cite{Rauer2020}. Kernels turn out to be extremely
convenient to encode physical constraints because they work within the
realm of linear algebra. Extending these properties to neural networks
and deep learning is more challenging, though the improved
expressivity has motivated active developments (\emph{vide infra}).

The lessons learned to build ML models in chemistry and materials
science largely transfer to soft matter and biomolecules, where
similar constraints on the representation prevail
\cite{gkeka2020machine}. Screening studies that make use of
kernel-based ML have become prominent, for instance in protein--ligand
binding, but many typically use experimental data
\cite{bartok2017machine}. Using MD, the relevant data-scale regimes
typically require a CG approach. For instance in drug--membrane
thermodynamics, CG simulations of $\sim 10^3$ systems led to
predictions for $1.3 \cdot 10^6$ molecules, thanks to the CG model's
reduction of CCS \cite{hoffmann2019controlled}. The predictions
satisfied thermodynamic relations observed on smaller data sets,
strongly suggesting robust generalization.  While this study was based
on a top-down CG model, systematic approaches like the variational
force-matching method bode elegantly well with the loss function of an
ML model. This has resulted in several studies, and in particular
efforts at addressing the challenging question of mapping many
atomistic configurations to a single CG geometry \cite{john2017many,
Wang2019, Wang2020, Scherer2020}. 

Several challenges still lie ahead for a more robust description of
condensed liquid-state systems. For instance, a (macro)molecule is
never isolated, but embedded in its environment, such that a
representation may benefit by incorporating the neighboring solvent's
degrees of freedom \cite{Pipolo2017}. The nature of the systems
naturally calls for the development of ML-based force fields that
incorporate long-range interactions \cite{Grisafi2019}, as well as
more particle types. We also point out the critical role of the
configurational aspect: a single geometry is not representative, but
rather should incorporate information about the underlying Boltzmann
distribution \cite{Rauer2020}. More than anything else, high-quality
ML models require extensive training data. Soft matter needs large,
homogeneous databases analogous to what has been developed from DFT
calculations for electronic properties, e.g., the QM9 database
\cite{ramakrishnan2014quantum}.

\subsubsection{Deep learning.}
The extraordinary results achieved with deep learning in so many
scientific and technological fields have to do with the added
expressivity of these models.  Using a neural-network architecture
that connects several layers of nodes, input and output can be mapped
to generalize surprisingly well \cite{Sejnowski2020}. Compared to the
above-mentioned regimes, the added expressivity of deep learning comes
at a price: they rely on an overwhelming number of parameters, and a
non-convex problem to solve. Practically this entails many more
training data points necessary to parametrize a model, typically in
the range $N_\textup{compounds} \gtrsim 10^6$. 

The benefits of deep learning are far reaching: notably for drug
discovery---though so far with data generated from experiments
\cite{Goh2017, Chen2018deepdrug}, we also outlined some of the
distinct conceptual advantages a deep-learning approach offers for
sampling both across conformations and compositions
(Section~\ref{sec:sampling:conf}). In terms of representing molecules,
the inclusion of symmetries is also an essential aspect, requiring
extensive methodological work \cite{thomas2018tensor,
kondor2018generalization}. They open the door to so-called
\emph{physics-informed neural networks}, which aim at a synergistic
combination of the two approaches to reduce the training data,
effectively regularizing in small data-scale regimes
\cite{Raissi2019}. Deep learning offers exciting opportunities: for
instance graph convolutional neural networks (CNNs) offer a physically
intuitive representation for molecules, where nodes and edges
represent atoms and bonds. Graph CNNs offer appealing features:
differentiable, more easily interpretable, and better performing than
commonly used molecular fingerprints \cite{duvenaud2015convolutional}.

Harnessing the full potential of deep-learning models puts stringent
requirement on the number of compounds, which severely restricts what
can be achieved in terms of screening studies. Few MD studies have
reached data-scale regimes amenable to deep learning, but impressive
first steps show much promise, such as the prediction of transfer free
energies in lipid membranes \cite{Bennett2020}. It offers a glance at
the use of MD-based studies to train deep-learning models across the
CCS of biomolecular and soft materials.

\section{Screening applications}
\label{sec:app}

The following describes a number of MD-based screening applications
for various soft-matter and biomolecular systems. We order the
applications roughly in the number of compounds screened, from low to
high, and grouped by topics when deemed fitting. Beyond the range of
screening sizes, some of these applications result from intense and
long-standing scientific activities. For those, the present review
cannot do justice to the breadth of these research topics, but will
hopefully stimulate the reader in diving into complementary readings.


\subsection{Exploring conformational space with swarms of trajectories}
\label{sec:app:swarms}
Far from a screening at high throughput, this first application
focuses on the study of \emph{individual} (macro)molecules. While
slightly deviating from the greater objective to screen across
compounds, the conceptual approach and implementation undertaken here
is relevant for our topic, as it provides innovative solutions to
exploring conformational space. 

The problem at heart involves the determination of kinetic properties
for systems exhibiting relevant processes at long time
scales---\emph{long} compared to what would be considered reasonably
achievable by a single trajectory on a supercomputer. Supercomputers
tackle ambitious simulations by means of CPU or GPU parallelization.
Unfortunately, not everything is easy to parallelize: While one can
easily segment a simulation box to treat smaller cells concurrently,
MD numerically integrates the equations of motion in a serial
fashion---it is difficult to parallelize time. Folding@Home tackled
the problem by introducing two complementary aspects: a conceptual
approach to circumvent the long-time-scale sampling problem, and a
platform to implement it \cite{Shirts2000}.

The dynamics of complex systems is typically dominated by free-energy
barriers: thermal fluctuations will lead a system to dwell in a
conformational basin (i.e., a local minimum), before being
spontaneously pushed over a barrier. Assuming single-exponential
kinetics with (unknown) rate $k$, the probability for the system to
cross the barrier at time $t$ is given by $P_1(t) = k\exp(-kt)$.
Rather than wait for a single trajectory to cross over once, let many
copies attempt it over a short time. In the case of $M$ simulations,
the probability for the first simulation to cross at the same time $t$
is now $P_M(t) = M k \exp(-M k t)$, exhibiting an effective rate that
is $M$ times faster. The pioneering work of Pande and coworkers
demonstrated the value of the approach: running multiple instances of
a short simulation boosts the chances of seeing early crossing events,
and sufficiently many occurrences allow them to estimate the rate $k$,
as illustrated on the folding of small peptides and polymers
\cite{Pande2002}.

The second breakthrough of the Folding@Home consortium was to
establish a distributed-computing platform, powered by idle CPU power
contributed by anonymous users over the internet \cite{Shirts2000}.
Running many short, uncoupled simulations meant that they did not need
to run on the same supercomputer. All simulation instances need no
communication, since they independently sample the same conformational
space. Practically this was simply realized by $M$ copies of the same
initial configuration (typically with different seeds and velocities),
since the stochasticity of the dynamical process will quickly lead to
diverging trajectories. 

One of the early examples of the Folding@Home project aimed at the
folding kinetics of two mutants of the designed, 23-residue-long
mini-protein BBA5 \cite{Snow2002}. With a mean folding time on the
order of $10$~$\mu$s, it is considered a fast-folding protein, yet
very much a challenging time-scale for an all-atom MD
simulation---especially at the time the research was conducted.
Following the above-mentioned reasoning for single-exponential
kinetics, they estimated that for such a folding timescale, roughly 10
out of 10,000 individual trajectories should fold after 10~ns. Using
an implicit-solvent united-atom model, they showed that an
impressively large number of short simulations yielded excellent
agreement with laser temperature-jump experiments. 

Folding@Home has made significant contributions in elucidating the
protein-folding problem in silico \cite{Dill2012, Lane2013}. Early
applications were then superseded with Markov state models, a more
robust memoryless master-equation treatment of the kinetics, pioneered
by No\'e, Pande, Chodera, Bowman, and others \cite{No2008, Pande2010,
Chodera2014, msm2014, Husic2018}. 

Moving away from protein folding, a more recent application of
distributed-computing platforms focused on protein--ligand binding.
Using their distributed-computing platform GPUGRID, De Fabritiis and
coworkers demonstrated the value of the approach for PMF calculations
for standard binding free energies \cite{Buch2011optimized}. Buch
\etal reported an impressive study of the enzyme-inhibitor complex
trypsin-benzamidine: they performed 495 unbiased MD simulations of the
unbound ligand for 100~ns each \cite{Buch2010, Buch2011complete}. They
sampled a variety of binding events, but also several \emph{pathways},
allowing them to robustly estimate both the binding free energy, as
well as the on and off binding rates. Extensions to the modeling of
protein--protein association kinetics form to date one of the most
impressive developments in this area~\cite{Plattner2017}.

Distributed-computing platforms have had a conceptual impact as to how
the community increasingly approaches MD simulations: from
handcrafted, individual instances to swarms of trajectories. The
associated need for automation paves the way for different kinds of
high-throughput MD simulations. Spawning MD trajectories has since
been extended to exploring uncharted regions of the free-energy
landscape using machine learning \cite{Chiavazzo2017}.


\subsection{Protein-ligand binding}
\label{sec:app:ligands}

The ever-growing penetration of computational chemistry in drug
discovery has experienced its shares of challenges
\cite{Jorgensen2004}. Like any complex engineering problem, the design
of a drug entails many considerations and complementary problems to
solve. From membrane penetration, to toxicity, to pharmacokinetic and
pharmacodynamic considerations, we focus here solely on the
determination of protein-ligand binding.

Basic structure-based drug-design methods typically assume rigid
drug--target structures: starting from a crystal structure or homology
modeling, a ligand is docked near the receptor's active site; the
molecular configuration is then used to estimate binding, often using
empirical scoring functions as a proxy. While this type of virtual
screening accommodates a large number of compounds, it models the
complex as mostly rigid. The lack of flexibility is an issue, given
the recognized role of the conformational ensemble in biomolecular
activity \cite{Boehr2009}. The field moved from a static lock-and-key
binding paradigm to more dynamic pictures, such as induced fit or
conformational selection. This emphasizes the need for physics-based
methods that model not only structural flexibility, but more broadly
the relevant emergent phenomena following binding \cite{de2016role}.

Beyond flexibility, an accurate account of the binding free energy is
desired. Free energies are \emph{ensemble} properties, making the
scoring of any individual configuration a conceptually peculiar
exercise. Several methods have been developed and tested over the
years---the drug-design field having explored many methodologies to
strike the right balance between accuracy and throughput: from
end-point methods to rigorous calculations derived from statistical
mechanics.

One prominent example of an end-point method combines MD simulations
on the bound and unbound configurations, using an implicit solvent and
a Poisson-Boltzmann surface area solvation term (MM-PBSA). Brown and
Muchmore applied MM-PBSA to a set of 308 ligands bound to one of three
protein receptors \cite{Brown2009}. The breadth and scope of the study
is laudable: moving toward a high-throughput MD scheme to extract free
energies of binding. The moderate correlation coefficients (Pearson
coefficient $R^2 = 0.5 - 0.7$) are unfortunately a testament to the
difficulties end-point methods display in reliably directing drug
discovery \cite{Jorgensen2009, Borhani2011}.

Alchemical transformations provide a rigorous framework to compute
binding free energies \cite{Mobley2012}. Though many methodologies
exist \cite{Chipot2002, Michel2010}, we mention one equilibrium
techniques that aims at calculating the free energy upon transforming
from state {\bf A} to {\bf B}: Free-energy perturbation, introduced by
Zwanzig \cite{Zwanzig1954}, relies on exponential averaging
\begin{equation}
    \label{eq:fep}
    \Delta G_{{\bf A} \rightarrow {\bf B}} = G_{\bf B} - G_{\bf A} = 
    - k_{\rm B}T \ln \left\langle \exp\left(-\frac{{\cal H}_{\rm B}({\bi r}) 
    - {\cal H}_{\rm A}({\bi r})}
    {k_{\rm B}T} \right)\right\rangle_{\bf A},
\end{equation}
where ${\bi r}$ denotes the system's particle coordinates, ${\cal
H}_{\rm A}$ is the Hamiltonian of state {\bf A}, and $\langle \cdot
\rangle_{\bf A}$ is an ensemble average at state point {\bf A}. 

Three decades ago, the pioneering study of Wong and McCammon presented
an alchemical transformation between benzamidine bound to the enzyme
trypsin \cite{Wong1986}. A fascinating review by Jorgensen describes
some of the successes of MD coupled with alchemical transformations to
advance the drug-discovery pipeline \cite{Jorgensen2009}. While the
generation of new scaffolds (i.e., entirely different structures) is
naturally sought, so-called hit-to-lead optimization---refinement of
the binding of a promising starting compound---is where alchemical
transformations really shine. There are two reasons for this: ($i$)
the computational expense of each alchemical transformation limits the
screening to relatively few compounds, thereby limiting the chances of
finding new scaffolds; and ($ii$) the interpolative nature of an
alchemical transformation (i.e., overlap in the conformational spaces,
see \ref{eq:fep}) leads to better convergence for similar molecules.

Alchemical transformations took a more systematic turn with the study
of Wang \etal \cite{Wang2015}. They reported relative free-energy
calculations at an all-atom level with explicit solvent for an
impressive 200 ligands. This feat was aided by the deployment of MD
simulations on graphics processing units (GPU), as well as a
streamlined procedure to prepare and run alchemical transformations.
Critically, they optimized a ``perturbation graph,'' which measures
the maximum common substructure between any pair of compounds
\cite{Liu2013}. The algorithm minimizes the number of alchemical
transformations, while accommodating for both multiple pathways to
estimate statistical error and the presence of closed cycles (which
ought to yield no free-energy difference). With a total of 330
perturbations, they reported a root-mean-squared error against
experiments of only 1.1~kcal/mol. More recent work has reported
alchemical transformations for up to several thousands of ligands
\cite{Abel2017}. Force-field improvements, from OPLS2.1 to OPLS3 and
OPLS3e have yielded systematic improvements in binding free energies
\cite{Harder2015, Roos2019}. 

Three decades of MD-based computational drug design have shown
impressive developments: not only in the sheer number of compounds
(from 1 to thousands reported in a single study), but more importantly
in the convergence of the calculations via significantly longer
simulation trajectories, and an overall improvement of the force
fields. The significant contributions of industrial actors is a
testament to both the pressing needs of the pharmaceutical industry
and the opportunities offered by physics-based MD methods.


\subsection{Solvation of small molecules}
\label{sec:app:solvation}

The free energy of solvation of small molecules is in many ways an
antechamber to protein-ligand binding: it consists of the free-energy
difference of transferring a small molecule from the gas into a
condensed-phase environment. Rather than a protein pocket, solvation
is performed in a bulk liquid. The homogeneity of the medium makes the
calculations easier to converge, typically allowing for broader
studies that may accommodate significantly more compounds.

The study of Jorgensen and Ravimohan pioneered alchemical
transformations by converting methanol into ethane
\cite{Jorgensen1985}. They applied free-energy perturbation (covered
in section~\ref{sec:app:ligands}) to compute the relative free-energy
difference in hydration---solvation in water---of the two compounds.
An alchemical transformation between these two similar molecules helps
the calculation: it only requires decoupling the hydroxyl group and
coupling a methyl in its stead.

Modeling solvation has had significant impact as a proxy for more
complex phenomena---a prime example being protein folding (some of
which was covered in section \ref{sec:app:swarms}). The
protein-folding problem was always strongly pushed by computer
simulations \cite{Dill2012}.  Huang \etal reported an insightful study
on hydrophobic solvation, they calculated the free energy of solvation
for hard-sphere solutes of various sizes \cite{Huang2001}. These
solutes, though not directly linked to any particular chemistry, aimed
at a better phenomenological understanding of possibly large
hydrophobic regions exposed to water, such as in protein folding. Of
particular interest was the systematic change in the solute size and
comparison of the asymptotics against theory. In the same vein, the
early 2000s witnessed intense activities in accurate calculations of
hydration and transfer free energies of (neutral) amino-acid
side-chain analogs \cite{Villa2002, MacCallum2003, Shirts2003,
Shirts2005}.

Mobley \etal reported hydration free energies for a set of 44 small,
neutral molecules \cite{Mobley2007}. A larger set of 239 small neutral
organic molecules was later tested against various force-field
parameters and charge models \cite{Shivakumar2009, Shivakumar2010,
Shivakumar2012}. In parallel, Mobley \etal released the FreeSolv
database, a set of 504 neutral small organic molecules, with
comparison against experiments \cite{Mobley2009}. Such studies have
led to the more routine incorporation of hydration free energies in
validating force fields \cite{Harder2015, Roos2019}. Scaling up,
Bennett \etal recently reported an impressive $15 \cdot 10^3$
water--cyclohexane transfer free-energy calculations from all-atom
molecular dynamics \cite{Bennett2020}.

Experimental free-energy datasets such as FreeSolv are useful because
they cover much of the diversity of small drug-like molecules,
although the small number of compounds necessarily limits how
representative they are. ML models of \emph{in silico} hydration free
energies trained on different datasets---both experimental and
combinatorially generated---did not appropriately generalize across
each other, highlighting biases in the chemical space covered
\cite{Rauer2020}. Still, the increased size and breadth of the spanned
chemical space allow researchers to identify systematic problems with
force-field parameters for classes of compounds. The same holds true
at the CG level: the automated Martini parametrization scheme for
small molecules facilitates the calculation of partitioning free
energies for several hundred molecules \cite{bereau2015automated}. It
helped identify systematic issues with certain chemical groups, such
as rings or halogens, which new versions of the force field aim at
correcting \cite{Souza2020}. 

With a growing number of computational techniques to compute free
energies, how can one compare their predictive accuracy in a fair way?
Nicholls \etal set up an informal blind-test study, comparing
different methodologies for 17 small molecules
\cite{nicholls2008predicting}. This was later formalized through the
SAMPL challenge \cite{Guthrie2009, sampl}. The blind tests consisted
of teams applying their method to compounds for which solvation free
energies are known but unpublished or relatively inaccessible. It
avoids the risks of tuning model parameters that would skew results to
seem artificially more favorable. SAMPL2 introduced an explanatory
section to gain insight in (disclosed) unexpected experimental results
\cite{Geballe2010}. Later challenges have since occurred and keep
helping benchmark and refine computational methods \cite{Mobley2014}.


\subsection{Ionic liquids}
\label{sec:app:ionliq}

Ionic liquids (ILs) are salts. They exhibit a melting point or
glass-transition temperature below $100^\circ$, while so-called
``room-temperature'' ILs remain liquid below $0^\circ$. ILs typically
exhibit good thermal stability, low vapor pressures, and are able to
dissolve many compounds. This makes ILs interesting solvents in
sustainable chemistry, with technological applications such as solvent
for biomolecules or catalysis \cite{armand2009ionic}. Critically, ILs
are also conductive, which makes them candidates for use in
electrochemical applications. In parallel, the combinatorics of
association of cation--anion pairs leads to an extraordinary number of
possible ILs. The combination of the breadth of chemical structures
available and the variety of properties of interest has motivated a
number of quantitative structure--property relationship modeling,
albeit so far mostly exclusively from experimental data
\cite{le2012quantitative}.

Computer simulations have played a significant role in better
understanding ILs. Maginn pointed out that interests in ILs rose
coincidentally with the advent of computer simulations, which have
proven increasingly capable of shedding light on complex fluids
\cite{Maginn2009}. The complex structural, thermodynamic, and
dynamical aspects, including behavior at interfaces, viscosity, and
dynamical heterogeneity motivated computational studies at various
scales, from quantum-mechanical calculations to classical atomistic to
coarse-grained modeling \cite{wang2007understanding,
lynden2007simulations, bhargava2008modelling, Maginn2009}. 

Turning to computational screening, Osti \etal reported an insightful
study aimed at probing ion interactions and transport in solvated ILs
\cite{Osti2016}. They fixed the IL cation--anion pair
(1-butyl-3-methyl-imidazolium bis(trifluoromethylsulfonyl)), but
screened across four organic solvents: acetonitrile (CH$_3$CN),
methanol (CH$_3$OH), tetrahydrofuran (C$_4$H$_8$O), and
dichloromethane (CH$_2$Cl$_2$). The potential of mean force of
separating a cation--anion pair suggested clear correlations between
the energetics of the interaction and solvent polarity: a larger
dipole moment is better able to screen ion--ion interactions, thereby
decreasing the free energy of solvation. This clear trend was mirrored
in the dynamics: ion diffusivity showed a linear increase against the
solvent dipole moment. The results were corroborated by quasi-elastic
neutron scattering experiments, overall offering clear
structure--property relationships.

A larger, follow-up screening yielded surprising results
\cite{thompson2019scalable}. Thompson \etal extended the set of
systems they studied, both in terms of IL--solvent mixtures (18
increments in the range 0.1--0.95 mass fraction) and solvent chemistry
(22 solvents including nitriles, alcohols, halocarbons, carbonyls, and
glymes) for a total of 396 state points. This study both further
confirmed a previously observed trend---IL mass fraction against IL
diffusivity---and uncovered a new one---solvent diffusivity against IL
diffusivity. Critically, they revisited the previously observed trend
by Osti \etal between IL diffusivity and solvent dipole moment
\cite{Osti2016}: the incorporation of more compounds indicated no
strong correlation across the entire data set. The authors hinted at
the role of complementary solvent order parameters to recover clear
trends. Combined, the two studies by Osti \etal and later Thompson
\etal illustrate a decisive aspect: the inference of
structure--property relationships hinges on a representative set of
chemical compounds.

\subsection{Silicate glasses}
\label{sec:app:silicglass}

Glasses---materials that have been cooled significantly but without
crystallizing---are known as structurally similar to but dynamically
very different from liquids \cite{paul1989chemistry}. Glassy materials
play a key role in many technological areas, motivating the
optimization of their mechanical properties, from hardness to fracture
strength to elastic properties \cite{Wondraczek2011}. Glasses embody
an overwhelming class of materials, when considering not only the
compositional aspects---potentially including a large number of
elements of the periodic table---but also its strong
out-of-equilibrium nature, meaning that the processing of the material
can easily lead to kinetic traps.

Yang \etal recently presented a high-throughput MD study of silicate
glasses, in an effort to predict their Young's modulus
\cite{Yang2019}. They covered the ternary diagram of calcium
aluminosilicate (CAS), CaO--Al$_2$O$_3$--SiO$_2$, by use of 231
compositions over the domain in 5\%-mol regular increments. The
authors ran MD simulations with tailored force fields
\cite{Bouhadja2013} using a melt-quench procedure to prepare the
configurations. All efforts were made at providing a consistent
system-preparation and simulation protocol throughout the
compositional space studied, but some limiting regimes required
specific treatments: ($i$) Higher initial melting temperature for
samples with high SiO$_2$ concentrations, due to their higher
glass-transition temperatures; and ($ii$) Faster cooling rate for
samples with high CaO concentrations, as they otherwise tend to
crystallize. These aspects illustrate the challenges faced by the need
for consistent protocols across large regions of
chemical/compositional space.

From the simulation data, they predicted the Young's modulus across
the compositional space using different statistical models. All their
approaches---from polynomial regression to various flavors of machine
learning---led to excellent results, indicative of both a dense
sampling of the compositional domain and a smooth mapping to the
target property. Interestingly, they showed that fitting models to
available experimental data ($\sim 100$ points) led to severe biases:
($i$) Clustering of the available data leaves large domains without
any training points; and ($ii$) Significant uncertainty and systematic
errors between experiments can lead to large variations. While the
latter aspect can be alleviated by means of adequate regularization,
the former recalls the ever-present dangers of extrapolation.


\subsection{Membrane proteins}
\label{sec:app:membprot}

Building up on the modeling of soluble proteins (see Section
\ref{sec:app:swarms}), membrane proteins form an important subset due
to their biochemical impact: they form roughly $25\,\%$ of all human
proteins \cite{Fagerberg2010} and half of current drug targets
\cite{Bakheet2009}.  Membrane proteins typically exert significantly
more complexity than their soluble counterparts. Transmembrane
proteins in particular---those that span the membrane bilayer---evolve
in a highly complex environment at the interface between the membrane
and the aqueous environment. This complex environment is compounded by
the large sizes that membrane proteins typically exhibit, often made
of numerous $\alpha$ helices or a prominent $\beta$ barrel. As a
result, the size and heterogeneity of membrane proteins have made them
challenging, not only for structure determination \cite{White2004,
Moraes2014}, but also for computer simulations \cite{Kandt2007,
Ayton2009, Chavent2016, Jefferies2020}.

The computational modeling of membrane proteins has benefitted heavily
from particle-based coarse-grained models. An all-atom treatment of a
protein and its surrounding lipid membrane remains to date a heroic
effort: protein folding happens over much longer time scales in the
membrane, due to the much larger correlation times exerted in the
bilayer. Peptide folding and insertion in a lipid membrane has been
reported at an atomistic level, although using an implicit-membrane
description, thereby speeding up the peptide dynamics in the membrane
environment \cite{Im2005}. Alternatively, coarse-grained models offer
an appealing way to study peptide folding and insertion in explicit
membranes, thereby offering the means to monitor how the peptide
perturbs membrane structure \cite{Bereau2014, Bereau2015}.

A coarse-grained description of membrane proteins does not only allow
to study folding and insertion for one of them, it can also be used to
study a larger number of systems. Sansom \etal presented more than a
decade ago an impressive protocol to automate the preparation of
transmembrane proteins \cite{sansom2008coarse}. Starting from
experimentally determined protein structures---typically deposited in
the Protein Data Bank (PDB) \cite{pdb, Berman2000}---these
macromolecules typically lack structural information about the aqueous
and membrane environments. Running MD simulations of a membrane
protein requires first to solvate it in both a lipid membrane and an
aqueous environment. Atomistic protocols typically start from
\emph{equilibrated} lipid bilayers and place a hole to incorporate the
macromolecule \cite{Jo2009}. Instead, the CG protocol of Sansom \etal
did not order the lipids in any way, but rather incorporated them as
an unstructured ``soup.'' The soup spontaneously rearranged into a
bilayer, thanks to self assembly and the speedy molecular diffusion at
the CG level. Other CG based schemes have been developed to ease and
automate the generation of complex lipid bilayers
\cite{Wassenaar2015insane} and the assembly of membrane-protein
multimers \cite{Wassenaar2015daft}. We note that the Martini-like CG
model does not allow for secondary or tertiary structure
reorganization, and is instead restrained around the crystal structure
\cite{Monticelli2008}. 

The pioneering database of Sansom \etal contained 91 membrane proteins
and was made available together with a web server to easily visualize
structural information \cite{sansom2008coarse}. Though no longer
available today, the Sansom group later released an expanded database
of membrane proteins: MemProtMD \cite{memprotmd}. Based on a more
sophisticated pipeline, the CG-based preparation protocol was amended
by a backmapping to atomistic resolution \cite{Stansfeld2015}. They
also more systematically imported structures from the PDB. The shear
size and incomplete data annotation of the PDB led them to design
structural descriptors to detect $\alpha$-helical and $\beta$-barrel
membrane proteins. An ensemble analysis across structures allowed them
to gain insight in the probabilities of occurrence of amino acid side
chains with respect to the depth in the bilayer. The MemProtMD
database and associated web server contains more than 3,500 PDB
entries \cite{Newport2018}. A systematic connection with other
databases brings in additional metadata to group structures according
to their constituent proteins and family. The network of protein
databases helps automatically annotate these structures with valuable
information.

Beyond the screening of membrane proteins themselves, cell membranes
embed these biomolecules in complex plasma membranes, made of a wide
diversity of compounds. Corradi \etal studied the protein--lipid
interactions for 10 membrane proteins embedded in a model plasma
membrane made of 60 lipid species \cite{Corradi2018}. The authors
identified clear ``lipid fingerprints:'' preferential association of
certain lipid species to parts of the protein. This study highlights
the combinatorial challenge involved, not only through the shear
sampling of each system, but the extreme compositional diversity at
hand. 

\subsection{Oligopeptide self assembly}
\label{sec:app:sapep}

The use of oligopeptides, consisting of a small number of residues, to
self assemble nanostructures offers the promise of tunable
supramolecular functionalities, yet with ease of preparation,
biocompatibility, and degradability \cite{Zelzer2010, Uhlig2014}. They
are proving viable contenders for applications in biomedicine and
nanotechnology \cite{Burroughes1990, Koss2018}. Various types of
nanostructures can be achieved, including fibers, tubes, and sheets
\cite{Hartgerink2001, Smith2008}. This diversity stems from the vast
combination of 20 natural amino acids into sequences.

In a series of studies, Frederix \etal have set up a systematic
MD-based virtual screening protocol to establish clear
structure--property relationships between the amino-acid sequence of
short peptides and self assembly under aqueous conditions. Using the
CG Martini force field, they first probed the ability to reproduce
structural features of the well-characterized diphenylalanine (FF)
peptide \cite{frederix2011virtual}. The aggregation of 1,600
dipeptides for $1.5\,\mu$s of simulation time (approximately
accounting for the acceleration due to coarse-graining) generated a
tubular nanostructure whose dimensions are in agreement with X-ray
diffraction analysis of crystallized FF nanotubes \cite{Grbitz2006}.
This indicated that despite structural limitations of the Martini
force field to model protein secondary structure, it could yield
reasonable self-assembling features. Beyond the final structure, the
simulations also helped understand the \emph{mechanism} of formation:
from an initial random placement to quick ordering into sheet-like
aggregates, to vesicle formation, and finally long hollow tubes.

Scaling up, Frederix \etal screened \emph{exhaustively} the space of
all possible $20^2 = 400$ dipeptide combinations
\cite{frederix2011virtual}. Although coarse-graining significantly
speeds up the simulations, the scope of the study led the researchers
to rapidly probe early determinants of aggregation. They followed the
self assembly of 300 dipeptides for 400\,ns. They scored the peptides'
aggregation propensity by means of the solvent-accessible surface
area, relative to the initial well-mixed configuration. The score was
in good qualitative agreement with experimentally resolved structures,
for the few sequences available. Though in need of atomistic
refinement, the study highlights how CG simulations can sketch the
mapping between sequence and self-assembled nanostructure.

A follow-up study aimed at the broader exploration of all tripeptides:
$20^3 = 8,000$ in total \cite{frederix2015exploring}. They sought
compounds that simultaneously favored aggregation propensity
\emph{and} hydrophilicity. While a priori contradicting requirements,
their results testify to the broad diversity of possible systems,
including subtle intermediates capable of displaying surprising
properties. Extending the dipeptide study, their
aggregation-propensity score was combined with the water--octanol
partitioning coefficient to measure hydrophobicity. They identified a
significant number of peptides that were not strongly hydrophobic, yet
exhibit aggregation. The screening confirmed and extended design rules
for the placement of specific amino acids in a particular position
\cite{Chiti2003, Marchesan2012}. This includes steric effects in the
placement of aromatic residues close to the N-terminus, but also
charged amino acids on positions 1 and 3 as an architecture for
intermolecular salt-bridge formation. Critically, their virtual
screening procedure led for the first time to the subsequent synthesis
and experimental characterization of tripeptides able to form
hydrogels at neutral pH.

More complex oligopeptides were considered more recently by Thurston
and Ferguson: a synthetic peptide--$\Pi$--peptide symmetric triblock
architecture of the form NXXX-$\Pi$-XXXN, where X are amino acids and
$\Pi$ is a conjugated aromatic core \cite{Thurston2018}. To limit the
space of candidates, they initially restricted their study to one of
two aromatic cores, naphthalenediimide and perylenediimide, and the
five amino acids A, F, G, I, and V were motivated by prior work.
Aiming at optoelectronic functionality, their design objective
targeted the stabilization of $\pi$--$\pi$ stacking between
neighboring oligopeptides, measuring the distance between aromatic
cores as a proxy for electronic delocalization. They relied on an
atomistic resolution with an implicit-solvent model to more
efficiently sample the conformational space. Both free energies of
dimerization and trimerization were calculated using enhanced-sampling
MD on 26 peptides. Intermediate values of the dimerization and
trimerization free energies led to the most favorable properties, as a
tradeoff between sufficient interaction strength to drive assembly,
yet little enough to avoid kinetic traps. A quantitative
structure--property relationship (QSPR) model was then trained on
these select peptides and a large set of 247 molecular descriptors,
based on the PaDEL software package \cite{Yap2010}. The authors
motivated their choice over more sophisticated machine learning
approaches both for its interpretability, as well as the dataset's
\emph{high-dimensional, low-sample size} regime. Further MD validation
confirmed the predictability of the QSPR for largely apolar
sequences---similar to the 26 training peptides---and proposed a new
sequence unstudied by experiment. While the QSPR lacked
transferability to strongly polar residues, the results indicate that
adding a limited set of MD simulations should be straightforward and
effective.

A wider study, also aiming at optimizing optoelectronic properties,
was recently reported by Shmilovich \etal
\cite{shmilovich2020discovery}. The synthetic architecture
DXXX-OPV3-XXXD used a three-repeat oligophenylenevinylene $\pi$ core,
for its ability to assemble into optically and electronically active
nanoaggregates \cite{Wall2012}. Compared to the study of Thurston and
Ferguson, the wider space of $20^3 = 8,000$ peptides was tackled by
two complementary strategies: ($i$) CG simulations using the Martini
force field, and ($ii$) a deep representational active learning
approach. Following the pioneering work of G\'omez-Bombarelli \etal
\cite{GmezBombarelli2018}, they projected the discrete sequence space
into a low-dimensional \emph{continuous} representation. A variational
autoencoder was used to train a latent-space embedding
\cite{kingma2013auto}, based on basic topological features of the CG
beads of the Martini model. They trained a Gaussian process regression
(GPR) on the latent-space embedding to predict the propensity of self
assembly, and used a Bayesian optimization to select the ``next best''
candidates to be simulated. Iterating over several generations of this
loop, they were able to converge the GPR model by only simulating
$2.3\,\%$ of the space of sequences. This computational design
platform, which aptly combines molecular simulations for compound
measurement and data-driven methods to efficiently sample the sequence
space, holds many promises for the virtual screening of biomolecular
and soft materials.

\subsection{Drug--membrane permeabilities}
\label{sec:app:drugmem}

One beloved application of biomolecular simulations is the cell
membrane. Though composed of a large variety of molecules, many are
phospholipids. These amphiphiles can spontaneously self assemble to
form large mesoscale structures, such as vesicles. This
compartmentalization of the cell can still allow for exchange of
(macro)molecules---either via active transport (biology), or passively
by simple diffusion (thermodynamics). This latter aspect can be
considered by the concentration gradient of a solute molecule, such as
a drug, across a soft interface between two aqueous environments.
Expressing this as a one-dimensional Smoluchowski equation along the
normal to the membrane, $z$, leads to the inhomogeneous
solubility-diffusion model \cite{diamond1974interpretation,
marrink1994simulation}. The resulting quantity is the permeability
coefficient, $P$, a flux that accounts for the heterogeneity of the
environment by integration over $z$ the energetics of crossing
together with the local diffusivity
\begin{equation}
    \label{eq:perm}
    P^{-1} = \int {\rm d}z \frac{\exp \left[ \beta G(z) \right]} 
    {D(z)}.
\end{equation}
In this equation, $\beta = 1/k_{\rm B}T$ is the inverse temperature,
$D(z)$ is the local diffusivity, and $G(z)$ is the potential of mean
force (PMF)---it is the free energy required to cross the interface as
a function of the order parameter $z$. Interestingly this quantity is
not readily accessible from current experimental techniques, leaving
computer simulations as the gold standard.

The use of enhanced-sampling techniques, such as umbrella sampling,
offer the means to compute the PMF at an atomistic resolution and
gather unprecedented insight \cite{orsi2010passive, swift2013back}.
Unsurprisingly the calculation of $G(z)$ is tremendously difficult to
converge: approximately $10^5$ CPU-hours is required for a small rigid
molecule crossing a single-component lipid membrane using
explicit-solvent atomistic models. This unfortunately limits an
atomistic throughput to $\sim 10$ molecules per study
\cite{carpenter2014method, lee2016simulation, bennion2017predicting,
tse2018link}. 

Here again, CG models allow for a significant step up in the number of
compounds that can be screened. Beyond the reduced representation
speeding up convergence of each simulation, the mapping to a Martini
representation easily leads to large numbers of compounds (Section
\ref{sec:sampling}). Menichetti \etal reported the PMFs of $4.6 \cdot
10^5$ small molecules in a one-component
1,2-dioleoyl-\emph{sn}-glycero-3-phosphocholine (DOPC) membrane
\cite{menichetti2017insilico}. This collection of compounds resulted
from the exhaustive screening of all CG small molecules made of one
and two neutral Martini beads, 14 and 105, respectively. The resulting
set of PMFs showed a strict variety, which could be accurately
correlated to the water/octanol partitioning of the solute---a bulk
quantity to relate to structural features at the membrane interface.
The mapping between chemistry and CG representations was established
by coarse-graining subsets of the GDB \cite{Fink2007}, keeping
compounds that mapped to one- and two-bead representations. 

A follow-up study extended the screening from PMFs to the permeability
coefficient (\ref{eq:perm}) \cite{menichetti2018drug}. The CG
simulations did not inform the diffusivity term (problematic due to
inconsistent accelerations of the CG dynamics \cite{Rudzinski2019}),
but were instead taken from atomistic simulations, indicating weak
dependence on the solute's chemistry \cite{carpenter2014method}. The
results showed excellent agreement with atomistic simulations and
correlation with experiments, despite the minimalistic modeling
approach. Permeability coefficients were predicted for $5.1 \cdot 10^5$
small organic molecules. Projecting the permeability surface onto two
physically motivated descriptors (hydrophobicity and acidity, i.e.,
p$K_{\rm a}$) highlighted the localization of key chemical groups, and
their influence on the target property. It also challenged earlier
phenomenological models of solute permeation
\cite{menichetti2019revisiting}.

A further scale up in the number of compounds ``simply'' comes down to
a broader screening toward larger CG representations: from one- and
two-bead constructs to more. The combinatorics of the Martini bead
types, while more favorable than atomically-detailed chemistry, still
grow exponentially: 14, 105, 1470, and 19\,306 for one- to four-bead
constructs---only considering linear chains. Instead of an exhaustive
account, Hoffmann \etal presented an importance-sampling scheme to
navigate the space of compounds \cite{hoffmann2019controlled}. A
Metropolis-chain Monte Carlo scheme was devised by daisy-chaining
compounds via alchemical transformations, and using the relative free
energy in the Metropolis criterion. This led to a large network of
compounds sampled, and the use of closed thermodynamic cycles allowed
for small corrections to the free energies. The space of compounds
that was \emph{not} sampled was subsequently predicted using a simple
kernel-based ML model. Some of the predictions were explicitly
validated, but all followed simple linear relationships between
transfer free energies that had been identified for the smaller
compounds \cite{menichetti2017insilico}---the thermodynamics of the
system acted as an ML physical constraint \emph{global to the compound
dataset}. Overall it boosted the prediction of transfer free energies
to $1.3 \cdot 10^6$ small organic molecules.

Extending the high-throughput CG framework, compound screening can be
used to better understand differential stabilization between lipid
domains, as a proxy for small molecules modulating complex
multi-component lipid membranes \cite{Centi2020}. The difference in
PMF minima between the relevant environments stands as a
computationally appealing proxy for large-scale simulations of
membrane reorganization. The results could identify families of
compounds that could induce membrane mixing or demixing. Compound
screening and their effect on membrane thermodynamics may help us
better understand the mechanism of action of certain anesthetics
\cite{Cornell2017}.

\section{Outlook}
\label{sec:outlook}

The path toward \emph{in silico} compound screening of biomaterials
and soft materials seems clear, but still contains a number of
important hurdles before reaching large data-scale regimes. Automating
the preparation, parametrization, and analysis of molecular dynamics
(MD) simulations is necessary to reach a high throughput, and has
largely embodied the scope of this review. The other critical aspect
is our capacity to run enough MD simulations, clearly the main
bottleneck. In this sense, coarse-grained (CG) modeling has an
important role to play: its ability to emulate a complex systems with
fewer degrees of freedom offers a significant scale-up in the context
of screening. The added capability to reduce the size of chemical
space seems to be a promising way to ease the analysis and extraction
of structure--property relationships.

Beyond statics, \emph{in silico} compound screening will likely hold
essential to target \emph{dynamical} properties, such as mean-first
passage times, folding and nucleation rates, or even aging dynamics.
To achieve this, force-field methods need to improve the modeling of
dynamics---a statement that holds at all scales, though in particular
at the CG level. The perspective to move toward non-equilibrium
systems will require the means to incorporate \emph{processing}
effects in materials, leading to structure--process--property
relationships. Getting there will be challenging: non-equilibrium
systems have no well-defined free-energy surface, and they critically
depend on how the system is prepared \cite{doi2013soft}.

Last, compound screening needs tighter integration with experiments.
This is not only in light of verifying the \emph{in silico}
predictions, but a collaborative procedure between simulations and
experiment that is poised to further accelerate soft-materials
discovery.

\ack
I thank Andrew L.~Ferguson and Joseph F.~Rudzinski for critical
reading of the manuscript. I am grateful to several colleagues and
collaborators for insightful discussions on topics pertaining to this
review, including Denis Andrienko, Andrew L.~Ferguson, Kiran
H.~Kanekal, Kurt Kremer, Anatole von Lilienfeld, Roberto Menichetti,
and Joseph F.~Rudzinski. Icons on Figures~\ref{fig:data_scale},
\ref{fig:hts_protocol}, \ref{fig:data_scale_cg} made by Becris,
Eucalyp, Freepik, and Monkik from \url{www.flaticon.com}.

\section*{References}
\bibliographystyle{unsrt}
\bibliography{biblio}

\begin{thebibliography}{100}

\bibitem{Ceder2013}
Gerbrand Ceder and Kristin Persson.
\newblock The stuff of dreams.
\newblock {\em Sci. Am.}, 309(6):36--40, November 2013.

\bibitem{jain2013commentary}
Anubhav Jain, Shyue~Ping Ong, Geoffroy Hautier, Wei Chen, William~Davidson
  Richards, Stephen Dacek, Shreyas Cholia, Dan Gunter, David Skinner, Gerbrand
  Ceder, and Kristin~A. Persson.
\newblock Commentary: The materials project: A materials genome approach to
  accelerating materials innovation.
\newblock {\em {APL} Mater.}, 1(1):011002, 2013.

\bibitem{curtarolo2013high}
Stefano Curtarolo, Gus~LW Hart, Marco~Buongiorno Nardelli, Natalio Mingo,
  Stefano Sanvito, and Ohad Levy.
\newblock The high-throughput highway to computational materials design.
\newblock {\em Nat. Mater.}, 12(3):191--201, 2013.

\bibitem{pyzer2015high}
Edward~O Pyzer-Knapp, Changwon Suh, Rafael G{\'o}mez-Bombarelli, Jorge
  Aguilera-Iparraguirre, and Al{\'a}n Aspuru-Guzik.
\newblock What is high-throughput virtual screening? a perspective from organic
  materials discovery.
\newblock {\em Annu. Rev. Mater. Res.}, 45:195--216, 2015.

\bibitem{jain2016computational}
Anubhav Jain, Yongwoo Shin, and Kristin~A Persson.
\newblock Computational predictions of energy materials using density
  functional theory.
\newblock {\em Nat. Rev. Mater.}, 1(1):15004, 2016.

\bibitem{ramprasad2017machine}
Rampi Ramprasad, Rohit Batra, Ghanshyam Pilania, Arun Mannodi-Kanakkithodi, and
  Chiho Kim.
\newblock Machine learning in materials informatics: recent applications and
  prospects.
\newblock {\em npj Comput. Mater.}, 3(1):54, 2017.

\bibitem{Tkatchenko2020}
Alexandre Tkatchenko.
\newblock Machine learning for chemical discovery.
\newblock {\em Nat. Comm.}, 11(1), August 2020.

\bibitem{Mishra2008}
K.P. Mishra, L.~Ganju, M.~Sairam, P.K. Banerjee, and R.C. Sawhney.
\newblock A review of high throughput technology for the screening of natural
  products.
\newblock {\em Biomed. Pharmacother.}, 62(2):94--98, February 2008.

\bibitem{Mayr2009}
Lorenz~M Mayr and Dejan Bojanic.
\newblock Novel trends in high-throughput screening.
\newblock {\em Curr. Opin. Pharmacol.}, 9(5):580--588, October 2009.

\bibitem{Macarron2011}
Ricardo Macarron, Martyn~N. Banks, Dejan Bojanic, David~J. Burns, Dragan~A.
  Cirovic, Tina Garyantes, Darren V.~S. Green, Robert~P. Hertzberg, William~P.
  Janzen, Jeff~W. Paslay, Ulrich Schopfer, and G.~Sitta Sittampalam.
\newblock Impact of high-throughput screening in biomedical research.
\newblock {\em Nat. Rev. Drug Discov.}, 10(3):188--195, March 2011.

\bibitem{Potyrailo2011}
Radislav Potyrailo, Krishna Rajan, Klaus Stoewe, Ichiro Takeuchi, Bret
  Chisholm, and Hubert Lam.
\newblock Combinatorial and high-throughput screening of materials libraries:
  Review of state of the art.
\newblock {\em {ACS} Comb. Sci.}, 13(6):579--633, August 2011.

\bibitem{Muster2011}
T.H. Muster, A.~Trinchi, T.A. Markley, D.~Lau, P.~Martin, A.~Bradbury,
  A.~Bendavid, and S.~Dligatch.
\newblock A review of high throughput and combinatorial electrochemistry.
\newblock {\em Electrochim. Acta}, 56(27):9679--9699, November 2011.

\bibitem{du2016microfluidics}
Guansheng Du, Qun Fang, and Jaap~MJ den Toonder.
\newblock Microfluidics for cell-based high throughput screening platforms—a
  review.
\newblock {\em Anal. Chim. Acta}, 903:36--50, 2016.

\bibitem{Dobson2004}
Christopher~M. Dobson.
\newblock Chemical space and biology.
\newblock {\em Nature}, 432(7019):824--828, December 2004.

\bibitem{Hert2009}
J{\'{e}}r{\^{o}}me Hert, John~J Irwin, Christian Laggner, Michael~J Keiser, and
  Brian~K Shoichet.
\newblock Quantifying biogenic bias in screening libraries.
\newblock {\em Nat. Chem. Biol.}, 5(7):479--483, May 2009.

\bibitem{Lin2018}
Arkadii Lin, Dragos Horvath, Valentina Afonina, Gilles Marcou, Jean-Louis
  Reymond, and Alexandre Varnek.
\newblock Mapping of the available chemical space versus the chemical universe
  of lead-like compounds.
\newblock {\em {Chem. Med. Chem.}}, 13(6):540--554, January 2018.

\bibitem{frederix2011virtual}
Pim~WJM Frederix, Rein~V Ulijn, Neil~T Hunt, and Tell Tuttle.
\newblock Virtual screening for dipeptide aggregation: Toward predictive tools
  for peptide self-assembly.
\newblock {\em J. Phys. Chem. Lett.}, 2(19):2380--2384, 2011.

\bibitem{Fink2007}
Tobias Fink and Jean-Louis Reymond.
\newblock {Virtual exploration of the chemical universe up to 11 atoms of C, N,
  O, F}.
\newblock {\em J. Chem. Inf. Model.}, 47(2):342--353, 2007.

\bibitem{Warmuth2003}
Manfred~K. Warmuth, Jun Liao, Gunnar R\"{a}tsch, Michael Mathieson, Santosh
  Putta, and Christian Lemmen.
\newblock Active learning with support vector machines in the drug discovery
  process.
\newblock {\em J. Chem. Inf. Comput. Sci.}, 43(2):667--673, February 2003.

\bibitem{Saal2013}
James~E. Saal, Scott Kirklin, Muratahan Aykol, Bryce Meredig, and C.~Wolverton.
\newblock Materials design and discovery with high-throughput density
  functional theory: The open quantum materials database ({OQMD}).
\newblock {\em {JOM}}, 65(11):1501--1509, September 2013.

\bibitem{Jain2016}
Anubhav Jain, Yongwoo Shin, and Kristin~A. Persson.
\newblock Computational predictions of energy materials using density
  functional theory.
\newblock {\em Nat. Rev. Mater.}, 1(1), January 2016.

\bibitem{Himanen2019}
Lauri Himanen, Amber Geurts, Adam~Stuart Foster, and Patrick Rinke.
\newblock Data-driven materials science: Status, challenges, and perspectives.
\newblock {\em Adv. Sci.}, 6(21):1900808, September 2019.

\bibitem{doi2013soft}
Masao Doi.
\newblock {\em Soft Matter Physics}.
\newblock Oxford University Press, 2013.

\bibitem{peter2010multiscale}
Christine Peter and Kurt Kremer.
\newblock Multiscale simulation of soft matter systems.
\newblock {\em Faraday Discuss.}, 144:9--24, 2010.

\bibitem{ferguson2017machine}
Andrew~L Ferguson.
\newblock Machine learning and data science in soft materials engineering.
\newblock {\em J. Condens. Matter Phys.}, 30(4):043002, 2017.

\bibitem{bereau2018data}
Tristan Bereau.
\newblock Data-driven methods in multiscale modeling of soft matter.
\newblock In W.~Andreoni and S.~Yip, editors, {\em Handbook of Materials
  Modeling: Methods: Theory and Modeling}, pages 1--12. Springer, 2018.

\bibitem{Jackson2019}
Nicholas~E Jackson, Michael~A Webb, and Juan~J de~Pablo.
\newblock Recent advances in machine learning towards multiscale soft materials
  design.
\newblock {\em Curr. Opin. Chem. Eng.}, 23:106--114, March 2019.

\bibitem{Greeley2006}
Jeff Greeley, Thomas~F. Jaramillo, Jacob Bonde, Ib~Chorkendorff, and Jens~K.
  N{\o}rskov.
\newblock Computational high-throughput screening of electrocatalytic materials
  for hydrogen evolution.
\newblock {\em Nat. Mater.}, 5(11):909--913, October 2006.

\bibitem{Simon2010}
Carl~G. Simon and Sheng Lin-Gibson.
\newblock Combinatorial and high-throughput screening of biomaterials.
\newblock {\em Adv. Mater.}, 23(3):369--387, September 2010.

\bibitem{Kitchen2004}
Douglas~B. Kitchen, H{\'{e}}l{\`{e}}ne Decornez, John~R. Furr, and J\"{u}rgen
  Bajorath.
\newblock Docking and scoring in virtual screening for drug discovery: methods
  and applications.
\newblock {\em Nat. Rev. Drug Discov.}, 3(11):935--949, November 2004.

\bibitem{Hachmann2011}
Johannes Hachmann, Roberto Olivares-Amaya, Sule Atahan-Evrenk, Carlos
  Amador-Bedolla, Roel~S. S{\'{a}}nchez-Carrera, Aryeh Gold-Parker, Leslie
  Vogt, Anna~M. Brockway, and Al{\'{a}}n Aspuru-Guzik.
\newblock The harvard clean energy project: Large-scale computational screening
  and design of organic photovoltaics on the world community grid.
\newblock {\em J. Phys. Chem. Lett.}, 2(17):2241--2251, August 2011.

\bibitem{Hellberg1987}
Sven Hellberg, Michael Sjoestroem, Bert Skagerberg, and Svante Wold.
\newblock Peptide quantitative structure-activity relationships, a multivariate
  approach.
\newblock {\em J. Med. Chem.}, 30(7):1126--1135, July 1987.

\bibitem{topliss2012quantitative}
John Topliss.
\newblock {\em Quantitative structure-activity relationships of drugs},
  volume~19.
\newblock Elsevier, 2012.

\bibitem{le2012quantitative}
Tu~Le, V~Chandana Epa, Frank~R Burden, and David~A Winkler.
\newblock Quantitative structure--property relationship modeling of diverse
  materials properties.
\newblock {\em Chem. Rev.}, 112(5):2889--2919, 2012.

\bibitem{tadmor2011modeling}
Ellad~B Tadmor and Ronald~E Miller.
\newblock {\em Modeling Materials: Continuum, Atomistic and Multiscale
  techniques}.
\newblock Cambridge University Press, 2011.

\bibitem{vanderGiessen2020}
Erik van~der Giessen, Peter~A Schultz, Nicolas Bertin, Vasily~V Bulatov, Wei
  Cai, G{\'{a}}bor Cs{\'{a}}nyi, Stephen~M Foiles, M~G~D Geers, Carlos
  Gonz{\'{a}}lez, Markus H\"{u}tter, Woo~Kyun Kim, Dennis~M Kochmann, Javier
  LLorca, Ann~E Mattsson, J\"{o}rg Rottler, Alexander Shluger, Ryan~B Sills,
  Ingo Steinbach, Alejandro Strachan, and Ellad~B Tadmor.
\newblock Roadmap on multiscale materials modeling.
\newblock {\em Modelling Simul. Mater. Sci. Eng.}, 28(4):043001, March 2020.

\bibitem{szabo2012modern}
Attila Szabo and Neil~S Ostlund.
\newblock {\em Modern Quantum Chemistry: Introduction to Advanced Electronic
  Structure Theory}.
\newblock Courier Corporation, 2012.

\bibitem{binder1995monte}
Kurt Binder.
\newblock {\em Monte Carlo and molecular dynamics simulations in polymer
  science}.
\newblock Oxford University Press, 1995.

\bibitem{frenkel2001understanding}
Daan Frenkel and Berend Smit.
\newblock {\em Understanding molecular simulation: from algorithms to
  applications}, volume~1.
\newblock Elsevier, 2001.

\bibitem{Karplus2002}
Martin Karplus and J.~Andrew McCammon.
\newblock Molecular dynamics simulations of biomolecules.
\newblock {\em Nat. Struct. Biol.}, 9(9):646--652, September 2002.

\bibitem{Hansson2002}
Tomas Hansson, Chris Oostenbrink, and WilfredF van Gunsteren.
\newblock Molecular dynamics simulations.
\newblock {\em Curr. Opin. Struct. Biol.}, 12(2):190--196, April 2002.

\bibitem{rapaport2004art}
Dennis~C Rapaport.
\newblock {\em The Art of Molecular Dynamics Simulation}.
\newblock Cambridge university press, 2004.

\bibitem{Mori1965}
Hazime Mori.
\newblock Transport, collective motion, and brownian motion.
\newblock {\em Prog. Theor. Phys.}, 33(3):423--455, March 1965.

\bibitem{Zwanzig1973}
Robert Zwanzig.
\newblock Nonlinear generalized langevin equations.
\newblock {\em J. Stat. Phys.}, 9(3):215--220, November 1973.

\bibitem{Tuckerman1992}
M.~Tuckerman, B.~J. Berne, and G.~J. Martyna.
\newblock Reversible multiple time scale molecular dynamics.
\newblock {\em J. Chem. Phys.}, 97(3):1990--2001, August 1992.

\bibitem{Gear2003}
C.~William Gear, James~M. Hyman, Panagiotis~G Kevrekidid, Ioannis~G.
  Kevrekidis, Olof Runborg, and Constantinos Theodoropoulos.
\newblock Equation-free, coarse-grained multiscale computation: Enabling
  mocroscopic simulators to perform system-level analysis.
\newblock {\em Commun. Math. Sci.}, 1(4):715--762, 2003.

\bibitem{Chodera2014}
John~D Chodera and Frank No{\'{e}}.
\newblock Markov state models of biomolecular conformational dynamics.
\newblock {\em Curr. Opin. Struct. Biol.}, 25:135--144, April 2014.

\bibitem{bereau2016research}
Tristan Bereau, Denis Andrienko, and Kurt Kremer.
\newblock Research update: Computational materials discovery in soft matter.
\newblock {\em {APL} Mater.}, 4(5):053101, 2016.

\bibitem{kaipio2006statistical}
Jari Kaipio and Erkki Somersalo.
\newblock {\em Statistical and Computational Inverse Problems}, volume 160.
\newblock Springer Science \& Business Media, 2006.

\bibitem{sherman2020inverse}
Zachary~M. Sherman, Michael~P. Howard, Beth~A. Lindquist, Ryan~B. Jadrich, and
  Thomas~M. Truskett.
\newblock Inverse methods for design of soft materials.
\newblock {\em J. Chem. Phys.}, 152(14):140902, April 2020.

\bibitem{Reymond2015}
Jean-Louis Reymond.
\newblock The chemical space project.
\newblock {\em Acc. Chem. Res.}, 48(3):722--730, February 2015.

\bibitem{stone2013theory}
Anthony Stone.
\newblock {\em The Theory of Intermolecular Forces}.
\newblock oUP oxford, 2013.

\bibitem{noid2013perspective}
W.~G. Noid.
\newblock Perspective: Coarse-grained models for biomolecular systems.
\newblock {\em J. Chem. Phys.}, 139(9):09B201\_1, 2013.

\bibitem{Inglfsson2013}
Helgi~I. Ing{\'{o}}lfsson, Cesar~A. Lopez, Jaakko~J. Uusitalo, Djurre~H.
  de~Jong, Srinivasa~M. Gopal, Xavier Periole, and Siewert~J. Marrink.
\newblock The power of coarse graining in biomolecular simulations.
\newblock {\em Wiley Interdiscip. Rev. Comput. Mol. Sci.}, 4(3):225--248,
  August 2013.

\bibitem{Hannon2013}
Adam~F. Hannon, Kevin~W. Gotrik, Caroline~A. Ross, and Alfredo Alexander-Katz.
\newblock Inverse design of topographical templates for directed self-assembly
  of block copolymers.
\newblock {\em {ACS} Macro Lett.}, 2(3):251--255, March 2013.

\bibitem{Miskin2015}
Marc~Z. Miskin, Gurdaman Khaira, Juan~J. de~Pablo, and Heinrich~M. Jaeger.
\newblock Turning statistical physics models into materials design engines.
\newblock {\em Proc. Natl. Acad. Sci. U.S.A.}, 113(1):34--39, December 2015.

\bibitem{Hormoz2011}
S.~Hormoz and M.~P. Brenner.
\newblock Design principles for self-assembly with short-range interactions.
\newblock {\em Proc. Natl. Acad. Sci. U.S.A.}, 108(13):5193--5198, March 2011.

\bibitem{vanAnders2015}
Greg van Anders, Daphne Klotsa, Andrew~S. Karas, Paul~M. Dodd, and Sharon~C.
  Glotzer.
\newblock Digital alchemy for materials design: Colloids and beyond.
\newblock {\em {ACS} Nano}, 9(10):9542--9553, October 2015.

\bibitem{Jain2014}
Avni Jain, Jonathan~A. Bollinger, and Thomas~M. Truskett.
\newblock Inverse methods for material design.
\newblock {\em {AIChE} Journal}, 60(8):2732--2740, May 2014.

\bibitem{Meng2010}
G.~Meng, N.~Arkus, M.~P. Brenner, and V.~N. Manoharan.
\newblock The free-energy landscape of clusters of attractive hard spheres.
\newblock {\em Science}, 327(5965):560--563, January 2010.

\bibitem{Bryngelson1995}
Joseph~D. Bryngelson, Jos{\'{e}}~Nelson Onuchic, Nicholas~D. Socci, and
  Peter~G. Wolynes.
\newblock Funnels, pathways, and the energy landscape of protein folding: A
  synthesis.
\newblock {\em Proteins: Struct., Funct., Bioinf.}, 21(3):167--195, March 1995.

\bibitem{Shakhnovich2006}
Eugene Shakhnovich.
\newblock Protein folding thermodynamics and dynamics:~ where physics,
  chemistry, and biology meet.
\newblock {\em Chem. Rev.}, 106(5):1559--1588, May 2006.

\bibitem{Dill2012}
K.~A. Dill and J.~L. MacCallum.
\newblock The protein-folding problem, 50 years on.
\newblock {\em Science}, 338(6110):1042--1046, November 2012.

\bibitem{Kuhlman2004}
Brian Kuhlman and David Baker.
\newblock Exploring folding free energy landscapes using computational protein
  design.
\newblock {\em Curr. Opin. Struct. Biol.}, 14(1):89--95, February 2004.

\bibitem{Jankowski2012}
Eric Jankowski and Sharon~C. Glotzer.
\newblock Screening and designing patchy particles for optimized self-assembly
  propensity through assembly pathway engineering.
\newblock {\em Soft Matter}, 8(10):2852, 2012.

\bibitem{menichetti2018drug}
Roberto Menichetti, Kiran~H Kanekal, and Tristan Bereau.
\newblock Drug--membrane permeability across chemical space.
\newblock {\em {ACS} Centr. Sci.}, 5(2):290--298, 2019.

\bibitem{serdyuk2017methods}
Igor~N Serdyuk, Nathan~R Zaccai, Joseph Zaccai, and Giuseppe Zaccai.
\newblock {\em Methods in Molecular Biophysics}.
\newblock Cambridge University Press, 2017.

\bibitem{maple1988derivation}
Jon~R Maple, Uri Dinur, and Arnold~T Hagler.
\newblock Derivation of force fields for molecular mechanics and dynamics from
  ab initio energy surfaces.
\newblock {\em Proc. Natl. Acad. Sci. U.S.A.}, 85(15):5350--5354, 1988.

\bibitem{halgren2001polarizable}
Thomas~A Halgren and Wolfgang Damm.
\newblock Polarizable force fields.
\newblock {\em Curr. Opin. Struct. Biol.}, 11(2):236--242, 2001.

\bibitem{wang2001biomolecular}
Wei Wang, Oreola Donini, Carolina~M Reyes, and Peter~A Kollman.
\newblock Biomolecular simulations: recent developments in force fields,
  simulations of enzyme catalysis, protein-ligand, protein-protein, and
  protein-nucleic acid noncovalent interactions.
\newblock {\em Annu. Rev. Bioph. Biom.}, 30(1):211--243, 2001.

\bibitem{ponder2003force}
Jay~W Ponder and David~A Case.
\newblock Force fields for protein simulations.
\newblock {\em Adv. Protein Chem.}, 66:27--85, 2003.

\bibitem{mackerell2004empirical}
Alexander~D Mackerell.
\newblock Empirical force fields for biological macromolecules: overview and
  issues.
\newblock {\em J. Comp. Chem.}, 25(13):1584--1604, 2004.

\bibitem{wang2017building}
Lee-Ping Wang, Keri~A McKiernan, Joseph Gomes, Kyle~A Beauchamp, Teresa
  Head-Gordon, Julia~E Rice, William~C Swope, Todd~J Mart{\'\i}nez, and Vijay~S
  Pande.
\newblock Building a more predictive protein force field: a systematic and
  reproducible route to amber-fb15.
\newblock {\em J. Phys. Chem. B}, 121(16):4023--4039, 2017.

\bibitem{halgren1992representation}
Thomas~A Halgren.
\newblock The representation of van der waals (vdw) interactions in molecular
  mechanics force fields: potential form, combination rules, and vdw
  parameters.
\newblock {\em J. Am. Chem. Soc.}, 114(20):7827--7843, 1992.

\bibitem{tkatchenko2012accurate}
Alexandre Tkatchenko, Robert~A DiStasio~Jr, Roberto Car, and Matthias
  Scheffler.
\newblock Accurate and efficient method for many-body van der waals
  interactions.
\newblock {\em Phys. Rev. Lett.}, 108(23):236402, 2012.

\bibitem{van2016beyond}
Mary~J Van~Vleet, Alston~J Misquitta, Anthony~J Stone, and Jordan~R Schmidt.
\newblock Beyond born--mayer: Improved models for short-range repulsion in ab
  initio force fields.
\newblock {\em J. Chem. Theory Comput.}, 12(8):3851--3870, 2016.

\bibitem{vanommeslaeghe2015charmm}
K.~Vanommeslaeghe and A.~D. MacKerell~Jr.
\newblock Charmm additive and polarizable force fields for biophysics and
  computer-aided drug design.
\newblock {\em Biochim. Biophys. Acta}, 1850(5):861--871, 2015.

\bibitem{Harder2015}
Edward Harder, Wolfgang Damm, Jon Maple, Chuanjie Wu, Mark Reboul, Jin~Yu
  Xiang, Lingle Wang, Dmitry Lupyan, Markus~K. Dahlgren, Jennifer~L. Knight,
  Joseph~W. Kaus, David~S. Cerutti, Goran Krilov, William~L. Jorgensen, Robert
  Abel, and Richard~A. Friesner.
\newblock {OPLS}3: A force field providing broad coverage of drug-like small
  molecules and proteins.
\newblock {\em J. Chem. Theory Comput.}, 12(1):281--296, December 2015.

\bibitem{Roos2019}
Katarina Roos, Chuanjie Wu, Wolfgang Damm, Mark Reboul, James~M. Stevenson,
  Chao Lu, Markus~K. Dahlgren, Sayan Mondal, Wei Chen, Lingle Wang, Robert
  Abel, Richard~A. Friesner, and Edward~D. Harder.
\newblock {OPLS}3e: Extending force field coverage for drug-like small
  molecules.
\newblock {\em J. Chem. Theory Comput.}, 15(3):1863--1874, February 2019.

\bibitem{Malde2011}
Alpeshkumar~K. Malde, Le~Zuo, Matthew Breeze, Martin Stroet, David Poger,
  Pramod~C. Nair, Chris Oostenbrink, and Alan~E. Mark.
\newblock An automated force field topology builder ({ATB}) and repository:
  Version 1.0.
\newblock {\em J. Chem. Theory Comput.}, 7(12):4026--4037, November 2011.

\bibitem{Wang2006automatic}
Junmei Wang, Wei Wang, Peter~A. Kollman, and David~A. Case.
\newblock Automatic atom type and bond type perception in molecular mechanical
  calculations.
\newblock {\em J. Mol. Graph. Model.}, 25(2):247--260, October 2006.

\bibitem{Mobley2018}
David~L. Mobley, Caitlin~C. Bannan, Andrea Rizzi, Christopher~I. Bayly, John~D.
  Chodera, Victoria~T. Lim, Nathan~M. Lim, Kyle~A. Beauchamp, David~R.
  Slochower, Michael~R. Shirts, Michael~K. Gilson, and Peter~K. Eastman.
\newblock Escaping atom types in force fields using direct chemical perception.
\newblock {\em J. Chem. Theory Comput.}, 14(11):6076--6092, October 2018.

\bibitem{rasmussen2004gaussian}
Carl~Edward Rasmussen.
\newblock Gaussian processes in machine learning.
\newblock In {\em Advanced Lectures on Machine Learning}, pages 63--71.
  Springer, 2004.

\bibitem{rupp2012fast}
Matthias Rupp, Alexandre Tkatchenko, Klaus-Robert M{\"u}ller, and O~Anatole
  Von~Lilienfeld.
\newblock Fast and accurate modeling of molecular atomization energies with
  machine learning.
\newblock {\em Phys. Rev. Lett.}, 108(5):058301, 2012.

\bibitem{Wilkins2019}
David~M. Wilkins, Andrea Grisafi, Yang Yang, Ka~Un Lao, Robert~A. DiStasio, and
  Michele Ceriotti.
\newblock Accurate molecular polarizabilities with coupled cluster theory and
  machine learning.
\newblock {\em Proc. Natl. Acad. Sci. U.S.A.}, 116(9):3401--3406, February
  2019.

\bibitem{bereau2015transferable}
Tristan Bereau, Denis Andrienko, and O~Anatole von Lilienfeld.
\newblock Transferable atomic multipole machine learning models for small
  organic molecules.
\newblock {\em J. Chem. Theory Comput.}, 11(7):3225--3233, 2015.

\bibitem{bereau2018non}
Tristan Bereau, Robert~A DiStasio~Jr, Alexandre Tkatchenko, and O~Anatole von
  Lilienfeld.
\newblock Non-covalent interactions across organic and biological subsets of
  chemical space: Physics-based potentials parametrized from machine learning.
\newblock {\em J. Chem. Phys.}, 147(24):241706, 2018.

\bibitem{li2017machine}
Ying Li, Hui Li, Frank~C Pickard~IV, Badri Narayanan, Fatih~G Sen, Maria~KY
  Chan, Subramanian~KRS Sankaranarayanan, Bernard~R Brooks, and Beno\^it Roux.
\newblock Machine learning force field parameters from ab initio data.
\newblock {\em J. Chem. Theory Comput.}, 13(9):4492--4503, 2017.

\bibitem{bartok2010gaussian}
Albert~P Bart{\'o}k, Mike~C Payne, Risi Kondor, and G{\'a}bor Cs{\'a}nyi.
\newblock Gaussian approximation potentials: The accuracy of quantum mechanics,
  without the electrons.
\newblock {\em Phys. Rev. Lett.}, 104(13):136403, 2010.

\bibitem{behler2016perspective}
J{\"o}rg Behler.
\newblock Perspective: Machine learning potentials for atomistic simulations.
\newblock {\em J. Chem. Phys.}, 145(17):170901, 2016.

\bibitem{chmiela2017machine}
Stefan Chmiela, Alexandre Tkatchenko, Huziel~E Sauceda, Igor Poltavsky,
  Kristof~T Sch{\"u}tt, and Klaus-Robert M{\"u}ller.
\newblock Machine learning of accurate energy-conserving molecular force
  fields.
\newblock {\em Sci. Adv.}, 3(5):e1603015, 2017.

\bibitem{Smith2017}
J.~S. Smith, O.~Isayev, and A.~E. Roitberg.
\newblock {ANI}-1: an extensible neural network potential with {DFT} accuracy
  at force field computational cost.
\newblock {\em Chemical Science}, 8(4):3192--3203, 2017.

\bibitem{grisafi2019incorporating}
Andrea Grisafi and Michele Ceriotti.
\newblock Incorporating long-range physics in atomic-scale machine learning.
\newblock {\em J. Chem. Phys.}, 151(20):204105, 2019.

\bibitem{marrink2013perspective}
Siewert~J Marrink and D~Peter Tieleman.
\newblock Perspective on the martini model.
\newblock {\em Chem. Soc. Rev.}, 42(16):6801--6822, 2013.

\bibitem{bereau2015automated}
Tristan Bereau and Kurt Kremer.
\newblock Automated parametrization of the coarse-grained martini force field
  for small organic molecules.
\newblock {\em J. Chem. Theory Comput.}, 11(6):2783--2791, 2015.

\bibitem{Kanekal2019}
Kiran~H. Kanekal and Tristan Bereau.
\newblock Resolution limit of data-driven coarse-grained models spanning
  chemical space.
\newblock {\em J. Chem. Phys.}, 151(16):164106, October 2019.

\bibitem{Rhle2009}
Victor R\"{u}hle, Christoph Junghans, Alexander Lukyanov, Kurt Kremer, and
  Denis Andrienko.
\newblock Versatile object-oriented toolkit for coarse-graining applications.
\newblock {\em J. Chem. Theory Comput.}, 5(12):3211--3223, November 2009.

\bibitem{Dunn2017}
Nicholas J.~H. Dunn, Kathryn~M. Lebold, Michael~R. DeLyser, Joseph~F.
  Rudzinski, and W.G. Noid.
\newblock {BOCS}: Bottom-up open-source coarse-graining software.
\newblock {\em J. Phys. Chem. B}, 122(13):3363--3377, December 2017.

\bibitem{Chakraborty2020}
Maghesree Chakraborty, Jinyu Xu, and Andrew~D. White.
\newblock Is preservation of symmetry necessary for coarse-graining?
\newblock {\em Phys. Chem. Chem. Phys.}, 22(26):14998--15005, 2020.

\bibitem{Foley2020}
Thomas~T. Foley, Katherine~M. Kidder, M.~Scott Shell, and W.~G. Noid.
\newblock Exploring the landscape of model representations.
\newblock {\em Proc. Natl. Acad. Sci. U.S.A.}, page 202000098, September 2020.

\bibitem{Martnez2009}
L.~Mart{\'{\i}}nez, R.~Andrade, E.~G. Birgin, and J.~M. Mart{\'{\i}}nez.
\newblock {PACKMOL}: A package for building initial configurations for
  molecular dynamics simulations.
\newblock {\em J. Comp. Chem.}, 30(13):2157--2164, October 2009.

\bibitem{polymermodeler}
Benjamin Haley, Nate Wilson, Chunyu Li, Andrea Arguelles, Eugenio Jaramillo,
  and Alejandro Strachan.
\newblock Polymer modeler, 2018.

\bibitem{Jo2008}
Sunhwan Jo, Taehoon Kim, Vidyashankara~G. Iyer, and Wonpil Im.
\newblock {CHARMM}-{GUI}: A web-based graphical user interface for {CHARMM}.
\newblock {\em J. Comp. Chem.}, 29(11):1859--1865, March 2008.

\bibitem{wassenaar2015computational}
Tsjerk~A Wassenaar, Helgi~I Ing{\o}lfsson, Rainer~A B{\"o}ckmann, D~Peter
  Tieleman, and Siewert~J Marrink.
\newblock Computational lipidomics with insane: A versatile tool for generating
  custom membranes for mol. simul.s.
\newblock {\em J. Chem. Theory Comput.}, 11(5):2144--2155, 2015.

\bibitem{newport2019memprotmd}
Thomas~D Newport, Mark S~P Sansom, and Phillip~J Stansfeld.
\newblock The memprotmd database: a resource for membrane-embedded protein
  structures and their lipid interactions.
\newblock {\em Nucleic Acids Res.}, 47(D1):D390--D397, 2019.

\bibitem{girard2019hoobas}
Martin Girard, Ali Ehlen, Anisha Shakya, Tristan Bereau, and Monica~Olvera
  de~la Cruz.
\newblock Hoobas: A highly object-oriented builder for molecular dynamics.
\newblock {\em Comput. Mater. Sci.}, 167:25--33, 2019.

\bibitem{Summers2020}
Andrew~Z. Summers, Justin~B. Gilmer, Christopher~R. Iacovella, Peter~T.
  Cummings, and Clare MCabe.
\newblock {MoSDeF}, a python framework enabling large-scale computational
  screening of soft matter: Application to chemistry-property relationships in
  lubricating monolayer films.
\newblock {\em J. Chem. Theory Comput.}, 16(3):1779--1793, January 2020.

\bibitem{neale2011statistical}
Chris Neale, WF~Drew Bennett, D~Peter Tieleman, and R{\'e}gis Pom{\`e}s.
\newblock Statistical convergence of equilibrium properties in simulations of
  molecular solutes embedded in lipid bilayers.
\newblock {\em J. Chem. Theory Comput.}, 7(12):4175--4188, 2011.

\bibitem{Shaw2010}
D.~E. Shaw, P.~Maragakis, K.~Lindorff-Larsen, S.~Piana, R.~O. Dror, M.~P.
  Eastwood, J.~A. Bank, J.~M. Jumper, J.~K. Salmon, Y.~Shan, and W.~Wriggers.
\newblock Atomic-level characterization of the structural dynamics of proteins.
\newblock {\em Science}, 330(6002):341--346, October 2010.

\bibitem{Abrams2013}
Cameron Abrams and Giovanni Bussi.
\newblock Enhanced sampling in molecular dynamics using metadynamics,
  replica-exchange, and temperature-acceleration.
\newblock {\em Entropy}, 16(1):163--199, December 2013.

\bibitem{Bernardi2015}
Rafael~C. Bernardi, Marcelo~C.R. Melo, and Klaus Schulten.
\newblock Enhanced sampling techniques in molecular dynamics simulations of
  biological systems.
\newblock {\em Biochim. Biophys. Acta}, 1850(5):872--877, May 2015.

\bibitem{Valsson2016}
Omar Valsson, Pratyush Tiwary, and Michele Parrinello.
\newblock Enhancing important fluctuations: Rare events and metadynamics from a
  conceptual viewpoint.
\newblock {\em Annu. Rev. Phys. Chem.}, 67(1):159--184, May 2016.

\bibitem{Camilloni2018}
Carlo Camilloni and Fabio Pietrucci.
\newblock Advanced simulation techniques for the thermodynamic and kinetic
  characterization of biological systems.
\newblock {\em Adv. Phys. X}, 3(1):1477531, January 2018.

\bibitem{kingma2013auto}
Diederik~P Kingma and Max Welling.
\newblock Auto-encoding variational bayes.
\newblock {\em arXiv preprint arXiv:1312.6114}, 2013.

\bibitem{Chen2018}
Wei Chen and Andrew~L. Ferguson.
\newblock Molecular enhanced sampling with autoencoders: On-the-fly collective
  variable discovery and accelerated free energy landscape exploration.
\newblock {\em J. Comp. Chem.}, 39(25):2079--2102, September 2018.

\bibitem{Sultan2018}
Mohammad~M. Sultan, Hannah~K. Wayment-Steele, and Vijay~S. Pande.
\newblock Transferable neural networks for enhanced sampling of protein
  dynamics.
\newblock {\em J. Chem. Theory Comput.}, 14(4):1887--1894, March 2018.

\bibitem{Wehmeyer2018}
Christoph Wehmeyer and Frank No{\'{e}}.
\newblock Time-lagged autoencoders: Deep learning of slow collective variables
  for molecular kinetics.
\newblock {\em J. Chem. Phys.}, 148(24):241703, June 2018.

\bibitem{Ribeiro2018}
Jo{\~{a}}o Marcelo~Lamim Ribeiro, Pablo Bravo, Yihang Wang, and Pratyush
  Tiwary.
\newblock Reweighted autoencoded variational bayes for enhanced sampling
  ({RAVE}).
\newblock {\em J. Chem. Phys.}, 149(7):072301, August 2018.

\bibitem{Chiavazzo2017}
Eliodoro Chiavazzo, Roberto Covino, Ronald~R. Coifman, C.~William Gear,
  Anastasia~S. Georgiou, Gerhard Hummer, and Ioannis~G. Kevrekidis.
\newblock Intrinsic map dynamics exploration for uncharted effective
  free-energy landscapes.
\newblock {\em Proc. Natl. Acad. Sci. U.S.A.}, 114(28):E5494--E5503, June 2017.

\bibitem{kukharenko2016using}
Oleksandra Kukharenko, Kevin Sawade, Jakob Steuer, and Christine Peter.
\newblock Using dimensionality reduction to systematically expand
  conformational sampling of intrinsically disordered peptides.
\newblock {\em J. Chem. Theory Comput.}, 12(10):4726--4734, 2016.

\bibitem{Ceriotti2011}
Michele Ceriotti, Gareth~A. Tribello, and Michele Parrinello.
\newblock Simplifying the representation of complex free-energy landscapes
  using sketch-map.
\newblock {\em Proc. Natl. Acad. Sci. U.S.A.}, 108(32):13023--13028, July 2011.

\bibitem{Lemke2019}
Tobias Lemke and Christine Peter.
\newblock {EncoderMap}: Dimensionality reduction and generation of molecule
  conformations.
\newblock {\em J. Chem. Theory Comput.}, 15(2):1209--1215, January 2019.

\bibitem{perez2015accelerating}
Alberto Perez, Justin~L MacCallum, and Ken~A Dill.
\newblock Accelerating molecular simulations of proteins using bayesian
  inference on weak information.
\newblock {\em Proc. Natl. Acad. Sci. U.S.A.}, 112(38):11846--11851, 2015.

\bibitem{Lipinski2004}
Christopher~A. Lipinski.
\newblock Lead- and drug-like compounds: the rule-of-five revolution.
\newblock {\em Drug Discov. Today Technol.}, 1(4):337--341, December 2004.

\bibitem{Ruddigkeit2012}
Lars Ruddigkeit, Ruud van Deursen, Lorenz~C. Blum, and Jean-Louis Reymond.
\newblock Enumeration of 166 billion organic small molecules in the chemical
  universe database {GDB}-17.
\newblock {\em J. Chem. Inf. Model.}, 52(11):2864--2875, November 2012.

\bibitem{ramakrishnan2014quantum}
Raghunathan Ramakrishnan, Pavlo~O Dral, Matthias Rupp, and O~Anatole
  Von~Lilienfeld.
\newblock Quantum chemistry structures and properties of 134 kilo molecules.
\newblock {\em Sci. Data}, 1:140022, 2014.

\bibitem{hoja2020qm7}
Johannes Hoja, Leonardo~Medrano Sandonas, Brian~G Ernst, Alvaro
  Vazquez-Mayagoitia, Robert~A DiStasio~Jr, and Alexandre Tkatchenko.
\newblock Qm7-x: A comprehensive dataset of quantum-mechanical properties
  spanning the chemical space of small organic molecules.
\newblock {\em arXiv preprint arXiv:2006.15139}, 2020.

\bibitem{Grest1986}
Gary~S. Grest and Kurt Kremer.
\newblock Molecular dynamics simulation for polymers in the presence of a heat
  bath.
\newblock {\em Phys. Rev. A}, 33(5):3628--3631, May 1986.

\bibitem{Kremer1990}
Kurt Kremer and Gary~S. Grest.
\newblock Dynamics of entangled linear polymer melts:{\hspace{0.167em}} a
  molecular-dynamics simulation.
\newblock {\em J. Chem. Phys.}, 92(8):5057--5086, April 1990.

\bibitem{Zhang2015}
Guojie Zhang, Torsten Stuehn, Kostas~Ch. Daoulas, and Kurt Kremer.
\newblock Communication: One size fits all: Equilibrating chemically different
  polymer liquids through universal long-wavelength description.
\newblock {\em J. Chem. Phys.}, 142(22):221102, June 2015.

\bibitem{Everaers2020}
Ralf Everaers, Hossein~Ali Karimi-Varzaneh, Frank Fleck, Nils Hojdis, and
  Carsten Svaneborg.
\newblock Kremer{\textendash}grest models for commodity polymer melts: Linking
  theory, experiment, and simulation at the kuhn scale.
\newblock {\em Macromolecules}, 53(6):1901--1916, March 2020.

\bibitem{automartini}
\texttt{auto\_martini} repository.
\newblock \url{https://github.com/tbereau/auto_martini}.

\bibitem{Virshup2013}
Aaron~M. Virshup, Julia Contreras-Garc{\'{\i}}a, Peter Wipf, Weitao Yang, and
  David~N. Beratan.
\newblock Stochastic voyages into uncharted chemical space produce a
  representative library of all possible drug-like compounds.
\newblock {\em J. Am. Chem. Soc.}, 135(19):7296--7303, May 2013.

\bibitem{Hoksza2014}
David Hoksza, Petr {\v{S}}koda, Milan Vor{\v{s}}il{\'{a}}k, and Daniel Svozil.
\newblock Molpher: a software framework for systematic chemical space
  exploration.
\newblock {\em J. Cheminformatics}, 6(1), March 2014.

\bibitem{joyce1992directed}
Gerald~F Joyce.
\newblock Directed molecular evolution.
\newblock {\em Sci. Am.}, 267(6):90--99, 1992.

\bibitem{Chowdhury2019}
Ratul Chowdhury and Costas~D. Maranas.
\newblock From directed evolution to computational enzyme
  engineering{\textemdash}a review.
\newblock {\em {AIChE} Journal}, 66(3), November 2019.

\bibitem{hoffmann2019controlled}
Christian Hoffmann, Roberto Menichetti, Kiran~H Kanekal, and Tristan Bereau.
\newblock Controlled exploration of chemical space by machine learning of
  coarse-grained representations.
\newblock {\em Phys. Rev. E}, 100:033302, 2019.

\bibitem{Wang2014}
Lee-Ping Wang, Alexey Titov, Robert McGibbon, Fang Liu, Vijay~S. Pande, and
  Todd~J. Mart{\'{\i}}nez.
\newblock Discovering chemistry with an ab initio nanoreactor.
\newblock {\em Nat. Chem.}, 6(12):1044--1048, November 2014.

\bibitem{Meisner2019}
Jan Meisner, Xiaolei Zhu, and Todd~J. Mart{\'{\i}}nez.
\newblock Computational discovery of the origins of life.
\newblock {\em {ACS} Cent. Sci.}, 5(9):1493--1495, September 2019.

\bibitem{Mobley2012}
David~L. Mobley and Pavel~V. Klimovich.
\newblock Perspective: Alchemical free energy calculations for drug discovery.
\newblock {\em J. Chem. Phys.}, 137(23):230901, December 2012.

\bibitem{AnatolevonLilienfeld2009}
O.~Anatole von Lilienfeld.
\newblock Accurate ab initio energy gradients in chemical compound space.
\newblock {\em J. Chem. Phys.}, 131(16):164102, October 2009.

\bibitem{Balawender2013}
Robert Balawender, Meressa~A. Welearegay, Micha{\l} Lesiuk, Frank~De Proft, and
  Paul Geerlings.
\newblock Exploring chemical space with the alchemical derivatives.
\newblock {\em J. Chem. Theory Comput.}, 9(12):5327--5340, November 2013.

\bibitem{toBaben2016}
M.~to~Baben, J.~O. Achenbach, and O.~A. von Lilienfeld.
\newblock Guiding ab initio calculations by alchemical derivatives.
\newblock {\em J. Chem. Phys.}, 144(10):104103, March 2016.

\bibitem{Wang2006}
Mingliang Wang, Xiangqian Hu, David~N. Beratan, and Weitao Yang.
\newblock Designing molecules by optimizing potentials.
\newblock {\em J. Am. Chem. Soc.}, 128(10):3228--3232, March 2006.

\bibitem{vonLilienfeld2005}
O.~Anatole von Lilienfeld, Roberto~D. Lins, and Ursula Rothlisberger.
\newblock Variational particle number approach for rational compound design.
\newblock {\em Phys. Rev. Lett.}, 95(15), October 2005.

\bibitem{GmezBombarelli2018}
Rafael G{\'{o}}mez-Bombarelli, Jennifer~N. Wei, David Duvenaud,
  Jos{\'{e}}~Miguel Hern{\'{a}}ndez-Lobato, Benjam{\'{\i}}n
  S{\'{a}}nchez-Lengeling, Dennis Sheberla, Jorge Aguilera-Iparraguirre,
  Timothy~D. Hirzel, Ryan~P. Adams, and Al{\'{a}}n Aspuru-Guzik.
\newblock Automatic chemical design using a data-driven continuous
  representation of molecules.
\newblock {\em {ACS} Cent. Sci.}, 4(2):268--276, January 2018.

\bibitem{shmilovich2020discovery}
Kirill Shmilovich, Rachael~A. Mansbach, Hythem Sidky, Olivia~E. Dunne,
  Sayak~Subhra Panda, John~D. Tovar, and Andrew~L. Ferguson.
\newblock Discovery of self-assembling $\pi$-conjugated peptides by active
  learning-directed coarse-grained mol. simul.
\newblock {\em J. Phys. Chem. B}, 124(19):3873--3891, March 2020.

\bibitem{Woelfle2011}
Michael Woelfle, Piero Olliaro, and Matthew~H. Todd.
\newblock Open science is a research accelerator.
\newblock {\em Nat. Chem.}, 3(10):745--748, September 2011.

\bibitem{Huan2016}
Tran~Doan Huan, Arun Mannodi-Kanakkithodi, Chiho Kim, Vinit Sharma, Ghanshyam
  Pilania, and Rampi Ramprasad.
\newblock A polymer dataset for accelerated property prediction and design.
\newblock {\em Sci. Data}, 3(1), March 2016.

\bibitem{Audus2017}
Debra~J. Audus and Juan~J. de~Pablo.
\newblock Polymer informatics: Opportunities and challenges.
\newblock {\em {ACS} Macro Lett.}, 6(10):1078--1082, September 2017.

\bibitem{Barnett2020}
J.~Wesley Barnett, Connor~R. Bilchak, Yiwen Wang, Brian~C. Benicewicz, Laura~A.
  Murdock, Tristan Bereau, and Sanat~K. Kumar.
\newblock Designing exceptional gas-separation polymer membranes using machine
  learning.
\newblock {\em Sci. Adv.}, 6(20):eaaz4301, May 2020.

\bibitem{zenodo_url}
{Zenodo}.
\newblock \url{https://zenodo.org}.

\bibitem{figshare}
{figshare}.
\newblock \url{https://figshare.com}.

\bibitem{osf}
{Open Science Framework}.
\newblock \url{https://osf.io}.

\bibitem{pdb}
{Protein Data Bank}.
\newblock \url{https://rcsb.org}.

\bibitem{hoffmann2020molecular}
Christian Hoffmann, Alessia Centi, Roberto Menichetti, and Tristan Bereau.
\newblock Molecular dynamics trajectories for 630 coarse-grained drug-membrane
  permeations.
\newblock {\em Sci. Data}, 7(1):1--7, 2020.

\bibitem{Wilkinson2016}
Mark~D. Wilkinson, Michel Dumontier, IJsbrand~Jan Aalbersberg, Gabrielle
  Appleton, Myles Axton, Arie Baak, Niklas Blomberg, Jan-Willem Boiten,
  Luiz~Bonino da~Silva~Santos, Philip~E. Bourne, Jildau Bouwman, Anthony~J.
  Brookes, Tim Clark, Merc{\`{e}} Crosas, Ingrid Dillo, Olivier Dumon, Scott
  Edmunds, Chris~T. Evelo, Richard Finkers, Alejandra Gonzalez-Beltran,
  Alasdair~J.G. Gray, Paul Groth, Carole Goble, Jeffrey~S. Grethe, Jaap
  Heringa, Peter~A.C 't~Hoen, Rob Hooft, Tobias Kuhn, Ruben Kok, Joost Kok,
  Scott~J. Lusher, Maryann~E. Martone, Albert Mons, Abel~L. Packer, Bengt
  Persson, Philippe Rocca-Serra, Marco Roos, Rene van Schaik, Susanna-Assunta
  Sansone, Erik Schultes, Thierry Sengstag, Ted Slater, George Strawn,
  Morris~A. Swertz, Mark Thompson, Johan van~der Lei, Erik van Mulligen, Jan
  Velterop, Andra Waagmeester, Peter Wittenburg, Katherine Wolstencroft, Jun
  Zhao, and Barend Mons.
\newblock The {FAIR} guiding principles for sci. data management and
  stewardship.
\newblock {\em Sci. Data}, 3(1), March 2016.

\bibitem{draxl2020big}
Claudia Draxl and Matthias Scheffler.
\newblock Big data-driven materials science and its fair data infrastructure.
\newblock {\em Handbook of Materials Modeling: Methods: Theory and Modeling},
  pages 49--73, 2020.

\bibitem{Blaiszik2016}
B.~Blaiszik, K.~Chard, J.~Pruyne, R.~Ananthakrishnan, S.~Tuecke, and I.~Foster.
\newblock The materials data facility: Data services to advance materials
  science research.
\newblock {\em {JOM}}, 68(8):2045--2052, July 2016.

\bibitem{Jain2013}
Anubhav Jain, Shyue~Ping Ong, Geoffroy Hautier, Wei Chen, William~Davidson
  Richards, Stephen Dacek, Shreyas Cholia, Dan Gunter, David Skinner, Gerbrand
  Ceder, and Kristin~A. Persson.
\newblock Commentary: The materials project: A materials genome approach to
  accelerating materials innovation.
\newblock {\em {APL} Mater.}, 1(1):011002, July 2013.

\bibitem{draxl2018nomad}
Claudia Draxl and Matthias Scheffler.
\newblock Nomad: The fair concept for big data-driven materials science.
\newblock {\em Mrs Bulletin}, 43(9):676--682, 2018.

\bibitem{Tadmor2011}
E.~B. Tadmor, R.~S. Elliott, J.~P. Sethna, R.~E. Miller, and C.~A. Becker.
\newblock The potential of atomistic simulations and the knowledgebase of
  interatomic models.
\newblock {\em {JOM}}, 63(7):17--17, July 2011.

\bibitem{Tadmor2013}
Ellad~B. Tadmor, Ryan~S. Elliott, Simon~R. Phillpot, and Susan~B. Sinnott.
\newblock {NSF} cyberinfrastructures: A new paradigm for advancing materials
  simulation.
\newblock {\em Curr. Opin. Solid State Mater. Sci.}, 17(6):298--304, December
  2013.

\bibitem{molssi}
{The Molecular Sciences Software Institute}.
\newblock \url{https://molssi.org}.

\bibitem{fairdi}
{FAIR-DI}.
\newblock \url{https://www.fair-di.eu}.

\bibitem{Lo2018}
Yu-Chen Lo, Stefano~E. Rensi, Wen Torng, and Russ~B. Altman.
\newblock Machine learning in chemoinformatics and drug discovery.
\newblock {\em Drug Discov. Today}, 23(8):1538--1546, August 2018.

\bibitem{swift2013back}
Robert~V Swift and Rommie~E Amaro.
\newblock Back to the future: can physical models of passive membrane
  permeability help reduce drug candidate attrition and move us beyond {QSPR}?
\newblock {\em Chem. Biol. Drug. Des.}, 81(1):61--71, 2013.

\bibitem{zhang2017machine}
Lu~Zhang, Jianjun Tan, Dan Han, and Hao Zhu.
\newblock From machine learning to deep learning: progress in machine
  intelligence for rational drug discovery.
\newblock {\em Drug Discov. Today}, 22(11):1680--1685, 2017.

\bibitem{Thurston2018}
Bryce~A. Thurston and Andrew~L. Ferguson.
\newblock Machine learning and molecular design of self-assembling -conjugated
  oligopeptides.
\newblock {\em Mol. Simul.}, 44(11):930--945, May 2018.

\bibitem{Brunton2016}
Steven~L. Brunton, Joshua~L. Proctor, and J.~Nathan Kutz.
\newblock Discovering governing equations from data by sparse identification of
  nonlinear dynamical systems.
\newblock {\em Proc. Natl. Acad. Sci. U.S.A.}, 113(15):3932--3937, March 2016.

\bibitem{Udrescu2020}
Silviu-Marian Udrescu and Max Tegmark.
\newblock {AI} feynman: A physics-inspired method for symbolic regression.
\newblock {\em Sci. Adv.}, 6(16):eaay2631, April 2020.

\bibitem{ghiringhelli2015big}
Luca~M Ghiringhelli, Jan Vybiral, Sergey~V Levchenko, Claudia Draxl, and
  Matthias Scheffler.
\newblock Big data of materials science: critical role of the descriptor.
\newblock {\em Phys. Rev. Lett.}, 114(10):105503, 2015.

\bibitem{ouyang2018sisso}
Runhai Ouyang, Stefano Curtarolo, Emre Ahmetcik, Matthias Scheffler, and Luca~M
  Ghiringhelli.
\newblock Sisso: A compressed-sensing method for identifying the best
  low-dimensional descriptor in an immensity of offered candidates.
\newblock {\em Phys. Rev. Mat.}, 2(8):083802, 2018.

\bibitem{goldsmith2017uncovering}
Bryan~R Goldsmith, Mario Boley, Jilles Vreeken, Matthias Scheffler, and Luca~M
  Ghiringhelli.
\newblock Uncovering structure-property relationships of materials by subgroup
  discovery.
\newblock {\em New J. Phys.}, 19(1):013031, 2017.

\bibitem{Ceriotti2019}
Michele Ceriotti.
\newblock Unsupervised machine learning in atomistic simulations, between
  predictions and understanding.
\newblock {\em J. Chem. Phys.}, 150(15):150901, April 2019.

\bibitem{faber2017prediction}
Felix~A Faber, Luke Hutchison, Bing Huang, Justin Gilmer, Samuel~S Schoenholz,
  George~E Dahl, Oriol Vinyals, Steven Kearnes, Patrick~F Riley, and O~Anatole
  Von~Lilienfeld.
\newblock Prediction errors of molecular machine learning models lower than
  hybrid dft error.
\newblock {\em J. Chem. Theory Comput.}, 13(11):5255--5264, 2017.

\bibitem{ramakrishnan2017machine}
Raghunathan Ramakrishnan and O.~Anatole von Lilienfeld.
\newblock {\em Machine Learning, Quantum Chemistry, and Chemical Space},
  chapter~5, pages 225--256.
\newblock John Wiley \& Sons, Ltd, 2017.

\bibitem{rasmussen2006gaussian}
Carl~Edward Rasmussen and Christopher~KI Williams.
\newblock {\em Gaussian processes for machine learning}, volume~1.
\newblock MIT press Cambridge, 2006.

\bibitem{bartok2017machine}
Albert~P Bart{\'o}k, Sandip De, Carl Poelking, Noam Bernstein, James~R Kermode,
  G{\'a}bor Cs{\'a}nyi, and Michele Ceriotti.
\newblock Machine learning unifies the modeling of materials and molecules.
\newblock {\em Sci. Adv.}, 3(12):e1701816, 2017.

\bibitem{faber2018alchemical}
Felix~A Faber, Anders~S Christensen, Bing Huang, and O~Anatole von Lilienfeld.
\newblock Alchemical and structural distribution based representation for
  universal quantum machine learning.
\newblock {\em J. Chem. Phys.}, 148(24):241717, 2018.

\bibitem{von2018quantum}
O~Anatole von Lilienfeld.
\newblock Quantum machine learning in chemical compound space.
\newblock {\em Angew. Chem. Int. Ed. Engl.}, 57(16):4164--4169, 2018.

\bibitem{Behler2007}
J\"{o}rg Behler and Michele Parrinello.
\newblock Generalized neural-network representation of high-dimensional
  potential-energy surfaces.
\newblock {\em Phys. Rev. Lett.}, 98(14), April 2007.

\bibitem{huang2016communication}
B~Huang and OA~von Lilienfeld.
\newblock Communication: Understanding molecular representations in machine
  learning: The role of uniqueness and target similarity.
\newblock {\em J. Chem. Phys.}, 145(16):161102--161102, 2016.

\bibitem{glielmo2017accurate}
Aldo Glielmo, Peter Sollich, and Alessandro De~Vita.
\newblock Accurate interatomic force fields via machine learning with covariant
  kernels.
\newblock {\em Phys. Rev. B}, 95(21):214302, 2017.

\bibitem{bartok_gaussian_2015}
Albert~P. Bartók and Gábor Csányi.
\newblock Gaussian approximation potentials: {A} brief tutorial introduction.
\newblock {\em Int. J. Quantum Chem.}, 115(16):1051--1057, 2015.

\bibitem{Scherer2020}
Christoph Scherer, Ren{\'{e}} Scheid, Denis Andrienko, and Tristan Bereau.
\newblock Kernel-based machine learning for efficient simulations of molecular
  liquids.
\newblock {\em J. Chem. Theory Comput.}, 16(5):3194--3204, April 2020.

\bibitem{Veit2020}
Max Veit, David~M. Wilkins, Yang Yang, Robert~A. DiStasio, and Michele
  Ceriotti.
\newblock Predicting molecular dipole moments by combining atomic partial
  charges and atomic dipoles.
\newblock {\em J. Chem. Phys.}, 153(2):024113, July 2020.

\bibitem{Rauer2020}
Clemens Rauer and Tristan Bereau.
\newblock Hydration free energies from kernel-based machine learning:
  Compound-database bias.
\newblock {\em J. Chem. Phys.}, 153(1):014101, July 2020.

\bibitem{gkeka2020machine}
Paraskevi Gkeka, Gabriel Stoltz, Amir~Barati Farimani, Zineb Belkacemi, Michele
  Ceriotti, John Chodera, Aaron~R Dinner, Andrew Ferguson, Jean-Bernard
  Maillet, Herv{\'e} Minoux, et~al.
\newblock Machine learning force fields and coarse-grained variables in
  molecular dynamics: application to materials and biological systems.
\newblock {\em arXiv preprint arXiv:2004.06950}, 2020.

\bibitem{john2017many}
ST~John and G{\'a}bor Cs{\'a}nyi.
\newblock Many-body coarse-grained interactions using gaussian approximation
  potentials.
\newblock {\em J. Phys. Chem. B}, 121(48):10934--10949, 2017.

\bibitem{Wang2019}
Jiang Wang, Simon Olsson, Christoph Wehmeyer, Adri{\`{a}} P{\'{e}}rez,
  Nicholas~E. Charron, Gianni de~Fabritiis, Frank No{\'{e}}, and Cecilia
  Clementi.
\newblock Machine learning of coarse-grained molecular dynamics force fields.
\newblock {\em {ACS} Cent. Sci.}, April 2019.

\bibitem{Wang2020}
Jiang Wang, Stefan Chmiela, Klaus-Robert M\"{u}ller, Frank No{\'{e}}, and
  Cecilia Clementi.
\newblock Ensemble learning of coarse-grained molecular dynamics force fields
  with a kernel approach.
\newblock {\em J. Chem. Phys.}, 152(19):194106, May 2020.

\bibitem{Pipolo2017}
S.~Pipolo, M.~Salanne, G.~Ferlat, S.~Klotz, A.{\hspace{0.167em}}M. Saitta, and
  F.~Pietrucci.
\newblock Navigating at will on the water phase diagram.
\newblock {\em Phys. Rev. Lett.}, 119(24), December 2017.

\bibitem{Grisafi2019}
Andrea Grisafi and Michele Ceriotti.
\newblock Incorporating long-range physics in atomic-scale machine learning.
\newblock {\em J. Chem. Phys.}, 151(20):204105, November 2019.

\bibitem{Sejnowski2020}
Terrence~J. Sejnowski.
\newblock The unreasonable effectiveness of deep learning in artificial
  intelligence.
\newblock {\em Proc. Natl. Acad. Sci. U.S.A.}, page 201907373, January 2020.

\bibitem{Goh2017}
Garrett~B. Goh, Nathan~O. Hodas, and Abhinav Vishnu.
\newblock Deep learning for computational chemistry.
\newblock {\em J. Comp. Chem.}, 38(16):1291--1307, March 2017.

\bibitem{Chen2018deepdrug}
Hongming Chen, Ola Engkvist, Yinhai Wang, Marcus Olivecrona, and Thomas
  Blaschke.
\newblock The rise of deep learning in drug discovery.
\newblock {\em Drug Discov. Today}, 23(6):1241--1250, June 2018.

\bibitem{thomas2018tensor}
Nathaniel Thomas, Tess Smidt, Steven Kearnes, Lusann Yang, Li~Li, Kai Kohlhoff,
  and Patrick Riley.
\newblock Tensor field networks: Rotation-and translation-equivariant neural
  networks for 3d point clouds.
\newblock {\em arXiv preprint arXiv:1802.08219}, 2018.

\bibitem{kondor2018generalization}
Risi Kondor and Shubhendu Trivedi.
\newblock On the generalization of equivariance and convolution in neural
  networks to the action of compact groups.
\newblock {\em arXiv preprint arXiv:1802.03690}, 2018.

\bibitem{Raissi2019}
M.~Raissi, P.~Perdikaris, and G.E. Karniadakis.
\newblock Physics-informed neural networks: A deep learning framework for
  solving forward and inverse problems involving nonlinear partial differential
  equations.
\newblock {\em J. Comput. Phys.}, 378:686--707, February 2019.

\bibitem{duvenaud2015convolutional}
David~K Duvenaud, Dougal Maclaurin, Jorge Iparraguirre, Rafael Bombarell,
  Timothy Hirzel, Al{\'a}n Aspuru-Guzik, and Ryan~P Adams.
\newblock Convolutional networks on graphs for learning molecular fingerprints.
\newblock In {\em Advances in neural information processing systems}, pages
  2224--2232, 2015.

\bibitem{Bennett2020}
W.F.~Drew Bennett, Stewart He, Camille~L Bilodeau, Derek Jones, Delin Sun,
  Hyojin Kim, Jonathan Allen, Felice~C Lightstone, and Helgi~I
  Ing{\'{o}}lfsson.
\newblock Predicting small molecule transfer free energies by combining
  molecular dynamics simulations and deep learning.
\newblock {\em J. Chem. Inf. Model.}, August 2020.

\bibitem{Shirts2000}
Michael Shirts and Vijay~S. Pande.
\newblock Screen savers of the world unite!
\newblock {\em Science}, 290(5498):1903--1904, 2000.

\bibitem{Pande2002}
Vijay~S. Pande, Ian Baker, Jarrod Chapman, Sidney~P. Elmer, Siraj Khaliq,
  Stefan~M. Larson, Young~Min Rhee, Michael~R. Shirts, Christopher~D. Snow,
  Eric~J. Sorin, and Bojan Zagrovic.
\newblock Atomistic protein folding simulations on the submillisecond time
  scale using worldwide distributed computing.
\newblock {\em Biopolymers}, 68(1):91--109, December 2002.

\bibitem{Snow2002}
Christopher~D. Snow, Houbi Nguyen, Vijay~S. Pande, and Martin Gruebele.
\newblock Absolute comparison of simulated and experimental protein-folding
  dynamics.
\newblock {\em Nature}, 420(6911):102--106, October 2002.

\bibitem{Lane2013}
Thomas~J Lane, Diwakar Shukla, Kyle~A Beauchamp, and Vijay~S Pande.
\newblock To milliseconds and beyond: challenges in the simulation of protein
  folding.
\newblock {\em Curr. Opin. Struct. Biol.}, 23(1):58--65, February 2013.

\bibitem{No2008}
Frank No{\'{e}}.
\newblock Probability distributions of molecular observables computed from
  markov models.
\newblock {\em J. Chem. Phys.}, 128(24):244103, June 2008.

\bibitem{Pande2010}
Vijay~S. Pande, Kyle Beauchamp, and Gregory~R. Bowman.
\newblock Everything you wanted to know about markov state models but were
  afraid to ask.
\newblock {\em Methods}, 52(1):99--105, September 2010.

\bibitem{msm2014}
Gregory~R. Bowman, Vijay~S. Pande, and Frank No{\'{e}}, editors.
\newblock {\em An Introduction to Markov State Models and Their Application to
  Long Timescale Mol. Simul.}
\newblock Springer Netherlands, 2014.

\bibitem{Husic2018}
Brooke~E. Husic and Vijay~S. Pande.
\newblock Markov state models: From an art to a science.
\newblock {\em J. Am. Chem. Soc.}, 140(7):2386--2396, February 2018.

\bibitem{Buch2011optimized}
Ignasi Buch, S.~Kashif Sadiq, and Gianni~De Fabritiis.
\newblock Optimized potential of mean force calculations for standard binding
  free energies.
\newblock {\em J. Chem. Theory Comput.}, 7(6):1765--1772, May 2011.

\bibitem{Buch2010}
I.~Buch, M.~J. Harvey, T.~Giorgino, D.~P. Anderson, and G.~De Fabritiis.
\newblock High-throughput all-atom molecular dynamics simulations using
  distributed computing.
\newblock {\em J. Chem. Inf. Model.}, 50(3):397--403, March 2010.

\bibitem{Buch2011complete}
I.~Buch, T.~Giorgino, and G.~De Fabritiis.
\newblock Complete reconstruction of an enzyme-inhibitor binding process by
  molecular dynamics simulations.
\newblock {\em Proc. Natl. Acad. Sci. U.S.A.}, 108(25):10184--10189, June 2011.

\bibitem{Plattner2017}
Nuria Plattner, Stefan Doerr, Gianni~De Fabritiis, and Frank No{\'{e}}.
\newblock Complete protein{\textendash}protein association kinetics in atomic
  detail revealed by molecular dynamics simulations and markov modelling.
\newblock {\em Nat. Chem.}, 9(10):1005--1011, June 2017.

\bibitem{Jorgensen2004}
W.~L. Jorgensen.
\newblock The many roles of computation in drug discovery.
\newblock {\em Science}, 303(5665):1813--1818, March 2004.

\bibitem{Boehr2009}
David~D Boehr, Ruth Nussinov, and Peter~E Wright.
\newblock The role of dynamic conformational ensembles in biomolecular
  recognition.
\newblock {\em Nat. Chem. Biol.}, 5(11):789--796, October 2009.

\bibitem{de2016role}
Marco De~Vivo, Matteo Masetti, Giovanni Bottegoni, and Andrea Cavalli.
\newblock Role of molecular dynamics and related methods in drug discovery.
\newblock {\em J. Med. Chem.}, 59(9):4035--4061, 2016.

\bibitem{Brown2009}
Scott~P. Brown and Steven~W. Muchmore.
\newblock Large-scale application of high-throughput molecular mechanics with
  poisson-boltzmann surface area for routine physics-based scoring of
  protein-ligand complexes.
\newblock {\em J. Med. Chem.}, 52(10):3159--3165, May 2009.

\bibitem{Jorgensen2009}
William~L. Jorgensen.
\newblock Efficient drug lead discovery and optimization.
\newblock {\em Acc. Chem. Res.}, 42(6):724--733, June 2009.

\bibitem{Borhani2011}
David~W. Borhani and David~E. Shaw.
\newblock The future of molecular dynamics simulations in drug discovery.
\newblock {\em J. Comput. Aided Mol. Des.}, 26(1):15--26, December 2011.

\bibitem{Chipot2002}
Christophe Chipot and David~A. Pearlman.
\newblock Free energy calculations. the long and winding gilded road.
\newblock {\em Mol. Simul.}, 28(1-2):1--12, January 2002.

\bibitem{Michel2010}
Julien Michel and Jonathan~W. Essex.
\newblock Prediction of protein{\textendash}ligand binding affinity by free
  energy simulations: assumptions, pitfalls and expectations.
\newblock {\em J. Comput. Aided Mol. Des.}, 24(8):639--658, May 2010.

\bibitem{Zwanzig1954}
Robert~W. Zwanzig.
\newblock High-temperature equation of state by a perturbation method. i.
  nonpolar gases.
\newblock {\em J. Chem. Phys.}, 22(8):1420--1426, August 1954.

\bibitem{Wong1986}
Chung~F. Wong and J.~Andrew. McCammon.
\newblock Dynamics and design of enzymes and inhibitors.
\newblock {\em J. Am. Chem. Soc.}, 108(13):3830--3832, June 1986.

\bibitem{Wang2015}
Lingle Wang, Yujie Wu, Yuqing Deng, Byungchan Kim, Levi Pierce, Goran Krilov,
  Dmitry Lupyan, Shaughnessy Robinson, Markus~K. Dahlgren, Jeremy Greenwood,
  Donna~L. Romero, Craig Masse, Jennifer~L. Knight, Thomas Steinbrecher, Thijs
  Beuming, Wolfgang Damm, Ed~Harder, Woody Sherman, Mark Brewer, Ron Wester,
  Mark Murcko, Leah Frye, Ramy Farid, Teng Lin, David~L. Mobley, William~L.
  Jorgensen, Bruce~J. Berne, Richard~A. Friesner, and Robert Abel.
\newblock Accurate and reliable prediction of relative ligand binding potency
  in prospective drug discovery by way of a modern free-energy calculation
  protocol and force field.
\newblock {\em J. Am. Chem. Soc.}, 137(7):2695--2703, February 2015.

\bibitem{Liu2013}
Shuai Liu, Yujie Wu, Teng Lin, Robert Abel, Jonathan~P. Redmann, Christopher~M.
  Summa, Vivian~R. Jaber, Nathan~M. Lim, and David~L. Mobley.
\newblock Lead optimization mapper: automating free energy calculations for
  lead optimization.
\newblock {\em J. Comput. Aided Mol. Des.}, 27(9):755--770, September 2013.

\bibitem{Abel2017}
Robert Abel, Lingle Wang, Edward~D. Harder, B.~J. Berne, and Richard~A.
  Friesner.
\newblock Advancing drug discovery through enhanced free energy calculations.
\newblock {\em Acc. Chem. Res.}, 50(7):1625--1632, July 2017.

\bibitem{Jorgensen1985}
William~L. Jorgensen and C.~Ravimohan.
\newblock Monte carlo simulation of differences in free energies of hydration.
\newblock {\em J. Chem. Phys.}, 83(6):3050--3054, September 1985.

\bibitem{Huang2001}
David~M. Huang, Phillip~L. Geissler, and David Chandler.
\newblock Scaling of hydrophobic solvation free energies.
\newblock {\em J. Phys. Chem. B}, 105(28):6704--6709, July 2001.

\bibitem{Villa2002}
Alessandra Villa and Alan~E. Mark.
\newblock Calculation of the free energy of solvation for neutral analogs of
  amino acid side chains.
\newblock {\em J. Comp. Chem.}, 23(5):548--553, April 2002.

\bibitem{MacCallum2003}
Justin~L. MacCallum and D.~Peter Tieleman.
\newblock Calculation of the water-cyclohexane transfer free energies of
  neutral amino acid side-chain analogs using the {OPLS} all-atom force field.
\newblock {\em J. Comp. Chem.}, 24(15):1930--1935, September 2003.

\bibitem{Shirts2003}
Michael~R. Shirts, Jed~W. Pitera, William~C. Swope, and Vijay~S. Pande.
\newblock Extremely precise free energy calculations of amino acid side chain
  analogs: Comparison of common molecular mechanics force fields for proteins.
\newblock {\em J. Chem. Phys.}, 119(11):5740--5761, September 2003.

\bibitem{Shirts2005}
Michael~R. Shirts and Vijay~S. Pande.
\newblock Solvation free energies of amino acid side chain analogs for common
  molecular mechanics water models.
\newblock {\em J. Chem. Phys.}, 122(13):134508, April 2005.

\bibitem{Mobley2007}
David~L. Mobley, {\'{E}}lise Dumont, John~D. Chodera, and Ken~A. Dill.
\newblock Comparison of charge models for fixed-charge force fields:~
  small-molecule hydration free energies in explicit solvent.
\newblock {\em J. Phys. Chem. B}, 111(9):2242--2254, March 2007.

\bibitem{Shivakumar2009}
Devleena Shivakumar, Yuqing Deng, and Beno{\^{\i}}t Roux.
\newblock Computations of absolute solvation free energies of small molecules
  using explicit and implicit solvent model.
\newblock {\em J. Chem. Theory Comput.}, 5(4):919--930, March 2009.

\bibitem{Shivakumar2010}
Devleena Shivakumar, Joshua Williams, Yujie Wu, Wolfgang Damm, John Shelley,
  and Woody Sherman.
\newblock Prediction of absolute solvation free energies using molecular
  dynamics free energy perturbation and the {OPLS} force field.
\newblock {\em J. Chem. Theory Comput.}, 6(5):1509--1519, April 2010.

\bibitem{Shivakumar2012}
Devleena Shivakumar, Edward Harder, Wolfgang Damm, Richard~A. Friesner, and
  Woody Sherman.
\newblock Improving the prediction of absolute solvation free energies using
  the next generation {OPLS} force field.
\newblock {\em J. Chem. Theory Comput.}, 8(8):2553--2558, July 2012.

\bibitem{Mobley2009}
David~L. Mobley, Christopher~I. Bayly, Matthew~D. Cooper, Michael~R. Shirts,
  and Ken~A. Dill.
\newblock Small molecule hydration free energies in explicit solvent: An
  extensive test of fixed-charge atomistic simulations.
\newblock {\em J. Chem. Theory Comput.}, 5(2):350--358, January 2009.

\bibitem{Souza2020}
Paulo C.~T. Souza, Sebastian Thallmair, Paolo Conflitti, Carlos
  Ram{\'{\i}}rez-Palacios, Riccardo Alessandri, Stefano Raniolo, Vittorio
  Limongelli, and Siewert~J. Marrink.
\newblock Protein{\textendash}ligand binding with the coarse-grained martini
  model.
\newblock {\em Nat. Comm.}, 11(1), July 2020.

\bibitem{nicholls2008predicting}
Anthony Nicholls, David~L Mobley, J~Peter Guthrie, John~D Chodera,
  Christopher~I Bayly, Matthew~D Cooper, and Vijay~S Pande.
\newblock Predicting small-molecule solvation free energies: an informal blind
  test for computational chemistry.
\newblock {\em J. Med. Chem.}, 51(4):769--779, 2008.

\bibitem{Guthrie2009}
J.~Peter Guthrie.
\newblock A blind challenge for computational solvation free energies:
  Introduction and overview.
\newblock {\em J. Phys. Chem. B}, 113(14):4501--4507, April 2009.

\bibitem{sampl}
{SAMPL Challenges}.
\newblock \url{https://samplchallenges.github.io}.

\bibitem{Geballe2010}
Matthew~T. Geballe, A.~Geoffrey Skillman, Anthony Nicholls, J.~Peter Guthrie,
  and Peter~J. Taylor.
\newblock The {SAMPL}2 blind prediction challenge: introduction and overview.
\newblock {\em J. Comput. Aided Mol. Des.}, 24(4):259--279, April 2010.

\bibitem{Mobley2014}
David~L. Mobley, Karisa~L. Wymer, Nathan~M. Lim, and J.~Peter Guthrie.
\newblock Blind prediction of solvation free energies from the {SAMPL}4
  challenge.
\newblock {\em J. Comput. Aided Mol. Des.}, 28(3):135--150, March 2014.

\bibitem{armand2009ionic}
Michel Armand, Frank Endres, Douglas~R MacFarlane, Hiroyuki Ohno, and Bruno
  Scrosati.
\newblock Ionic-liquid materials for the electrochemical challenges of the
  future.
\newblock {\em Nat. Mater.}, 8(8):621--629, 2009.

\bibitem{Maginn2009}
E~J Maginn.
\newblock Molecular simulation of ionic liquids: current status and future
  opportunities.
\newblock {\em J. Phys. Condens. Matter.}, 21(37):373101, August 2009.

\bibitem{wang2007understanding}
Yanting Wang, WEI Jiang, Tianying Yan, and Gregory~A Voth.
\newblock Understanding ionic liquids through atomistic and coarse-grained
  molecular dynamics simulations.
\newblock {\em Acc. Chem. Res.}, 40(11):1193--1199, 2007.

\bibitem{lynden2007simulations}
Ruth~M Lynden-Bell, Mario~G Del~Popolo, Tristan~GA Youngs, Jorge Kohanoff,
  Christof~G Hanke, Jason~B Harper, and Carlos~C Pinilla.
\newblock Simulations of ionic liquids, solutions, and surfaces.
\newblock {\em Acc. Chem. Res.}, 40(11):1138--1145, 2007.

\bibitem{bhargava2008modelling}
BL~Bhargava, Sundaram Balasubramanian, and Michael~L Klein.
\newblock Modelling room temperature ionic liquids.
\newblock {\em Chem. Commun.}, 29:3339--3351, 2008.

\bibitem{Osti2016}
Naresh~C. Osti, Katherine L.~Van Aken, Matthew~W. Thompson, Felix Tiet,
  De~en~Jiang, Peter~T. Cummings, Yury Gogotsi, and Eugene Mamontov.
\newblock Solvent polarity governs ion interactions and transport in a solvated
  room-temperature ionic liquid.
\newblock {\em J. Phys. Chem. Lett.}, 8(1):167--171, December 2016.

\bibitem{thompson2019scalable}
Matthew~W Thompson, Ray Matsumoto, Robert~L Sacci, Nicolette~C Sanders, and
  Peter~T Cummings.
\newblock Scalable screening of soft matter: A case study of mixtures of ionic
  liquids and organic solvents.
\newblock {\em J. Phys. Chem. B}, 123(6):1340--1347, 2019.

\bibitem{paul1989chemistry}
Amal Paul.
\newblock {\em Chemistry of glasses}.
\newblock Springer Science \& Business Media, 1989.

\bibitem{Wondraczek2011}
Lothar Wondraczek, John~C. Mauro, J\"{u}rgen Eckert, Uta K\"{u}hn, J\"{u}rgen
  Horbach, Joachim Deubener, and Tanguy Rouxel.
\newblock Towards ultrastrong glasses.
\newblock {\em Adv. Mater.}, 23(39):4578--4586, September 2011.

\bibitem{Yang2019}
Kai Yang, Xinyi Xu, Benjamin Yang, Brian Cook, Herbert Ramos, N.~M.~Anoop
  Krishnan, Morten~M. Smedskjaer, Christian Hoover, and Mathieu Bauchy.
\newblock Predicting the {Young}'s modulus of silicate glasses using
  high-throughput molecular dynamics simulations and machine learning.
\newblock {\em Sci. Rep.}, 9(1), June 2019.

\bibitem{Bouhadja2013}
M.~Bouhadja, N.~Jakse, and A.~Pasturel.
\newblock Structural and dynamic properties of calcium aluminosilicate melts: A
  molecular dynamics study.
\newblock {\em J. Chem. Phys.}, 138(22):224510, June 2013.

\bibitem{Fagerberg2010}
Linn Fagerberg, Kalle Jonasson, Gunnar von Heijne, Mathias Uhl{\'{e}}n, and
  Lisa Berglund.
\newblock Prediction of the human membrane proteome.
\newblock {\em Proteomics}, 10(6):1141--1149, February 2010.

\bibitem{Bakheet2009}
Tala~M. Bakheet and Andrew~J. Doig.
\newblock Properties and identification of human protein drug targets.
\newblock {\em Bioinformatics}, 25(4):451--457, January 2009.

\bibitem{White2004}
Stephen~H. White.
\newblock The progress of membrane protein structure determination.
\newblock {\em Protein Sci.}, 13(7):1948--1949, July 2004.

\bibitem{Moraes2014}
Isabel Moraes, Gwyndaf Evans, Juan Sanchez-Weatherby, Simon Newstead, and
  Patrick D.~Shaw Stewart.
\newblock Membrane protein structure determination {\textemdash} the next
  generation.
\newblock {\em Biochim. Biophys. Acta Biomembranes}, 1838(1):78--87, January
  2014.

\bibitem{Kandt2007}
Christian Kandt, Walter~L. Ash, and D.~Peter Tieleman.
\newblock Setting up and running molecular dynamics simulations of membrane
  proteins.
\newblock {\em Methods}, 41(4):475--488, April 2007.

\bibitem{Ayton2009}
Gary~S Ayton and Gregory~A Voth.
\newblock Systematic multiscale simulation of membrane protein systems.
\newblock {\em Curr. Opin. Struct. Biol.}, 19(2):138--144, April 2009.

\bibitem{Chavent2016}
Matthieu Chavent, Anna~L Duncan, and Mark~SP Sansom.
\newblock Molecular dynamics simulations of membrane proteins and their
  interactions: from nanoscale to mesoscale.
\newblock {\em Curr. Opin. Struct. Biol.}, 40:8--16, October 2016.

\bibitem{Jefferies2020}
Damien Jefferies and Syma Khalid.
\newblock Atomistic and coarse-grained simulations of membrane proteins: A
  practical guide.
\newblock {\em Methods}, February 2020.

\bibitem{Im2005}
W.~Im and C.~L. Brooks.
\newblock Interfacial folding and membrane insertion of designed peptides
  studied by molecular dynamics simulations.
\newblock {\em Proc. Natl. Acad. Sci. U.S.A.}, 102(19):6771--6776, April 2005.

\bibitem{Bereau2014}
Tristan Bereau and Markus Deserno.
\newblock Enhanced sampling of coarse-grained transmembrane-peptide structure
  formation from hydrogen-bond replica exchange.
\newblock {\em J. Memb. Biol.}, 248(3):395--405, October 2014.

\bibitem{Bereau2015}
Tristan Bereau, W.~F.~Drew Bennett, Jim Pfaendtner, Markus Deserno, and Mikko
  Karttunen.
\newblock Folding and insertion thermodynamics of the transmembrane {WALP}
  peptide.
\newblock {\em J. Chem. Phys.}, 143(24):243127, December 2015.

\bibitem{sansom2008coarse}
MS~Sansom, KA~Scott, and PJ~Bond.
\newblock Coarse-grained simulation: a high-throughput computational approach
  to membrane proteins.
\newblock {\em Biochem. Soc. Trans.}, 36(Pt 1), 2008.

\bibitem{Berman2000}
H.~M. Berman.
\newblock The protein data bank.
\newblock {\em Nucleic Acids Res.}, 28(1):235--242, January 2000.

\bibitem{Jo2009}
Sunhwan Jo, Joseph~B. Lim, Jeffery~B. Klauda, and Wonpil Im.
\newblock {CHARMM}-{GUI} membrane builder for mixed bilayers and its
  application to yeast membranes.
\newblock {\em Biophys. J.}, 97(1):50--58, July 2009.

\bibitem{Wassenaar2015insane}
Tsjerk~A. Wassenaar, Helgi~I. Ing{\'{o}}lfsson, Rainer~A. B\"{o}ckmann,
  D.~Peter Tieleman, and Siewert~J. Marrink.
\newblock Computational lipidomics with insane: A versatile tool for generating
  custom membranes for mol. simul.s.
\newblock {\em J. Chem. Theory Comput.}, 11(5):2144--2155, April 2015.

\bibitem{Wassenaar2015daft}
Tsjerk~A. Wassenaar, Kristyna Pluhackova, Anastassiia Moussatova, Durba
  Sengupta, Siewert~J. Marrink, D.~Peter Tieleman, and Rainer~A. B\"{o}ckmann.
\newblock High-throughput simulations of dimer and trimer assembly of membrane
  proteins. the {DAFT} approach.
\newblock {\em J. Chem. Theory Comput.}, 11(5):2278--2291, March 2015.

\bibitem{Monticelli2008}
Luca Monticelli, Senthil~K. Kandasamy, Xavier Periole, Ronald~G. Larson,
  D.~Peter Tieleman, and Siewert-Jan Marrink.
\newblock The {MARTINI} coarse-grained force field: Extension to proteins.
\newblock {\em J. Chem. Theory Comput.}, 4(5):819--834, April 2008.

\bibitem{memprotmd}
{MemProtMD} database.
\newblock \url{http://memprotmd.bioch.ox.ac.uk/home/}.

\bibitem{Stansfeld2015}
Phillip~J. Stansfeld, Joseph~E. Goose, Martin Caffrey, Elisabeth~P. Carpenter,
  Joanne~L. Parker, Simon Newstead, and Mark~S.P. Sansom.
\newblock {MemProtMD}: Automated insertion of membrane protein structures into
  explicit lipid membranes.
\newblock {\em Structure}, 23(7):1350--1361, July 2015.

\bibitem{Newport2018}
Thomas~D Newport, Mark S~P Sansom, and Phillip~J Stansfeld.
\newblock The {MemProtMD} database: a resource for membrane-embedded protein
  structures and their lipid interactions.
\newblock {\em Nucleic Acids Res.}, 47(D1):D390--D397, November 2018.

\bibitem{Corradi2018}
Valentina Corradi, Eduardo Mendez-Villuendas, Helgi~I. Ing{\'{o}}lfsson, Ruo-Xu
  Gu, Iwona Siuda, Manuel~N. Melo, Anastassiia Moussatova, Lucien~J.
  DeGagn{\'{e}}, Besian~I. Sejdiu, Gurpreet Singh, Tsjerk~A. Wassenaar,
  Karelia~Delgado Magnero, Siewert~J. Marrink, and D.~Peter Tieleman.
\newblock Lipid{\textendash}protein interactions are unique fingerprints for
  membrane proteins.
\newblock {\em {ACS} Cent. Sci.}, 4(6):709--717, June 2018.

\bibitem{Zelzer2010}
Mischa Zelzer and Rein~V. Ulijn.
\newblock Next-generation peptide nanomaterials: molecular networks, interfaces
  and supramolecular functionality.
\newblock {\em Chem. Soc. Rev.}, 39(9):3351, 2010.

\bibitem{Uhlig2014}
Thomas Uhlig, Themis Kyprianou, Filippo~Giancarlo Martinelli, Carlo~Alberto
  Oppici, Dave Heiligers, Diederik Hills, Xavier~Ribes Calvo, and Peter
  Verhaert.
\newblock The emergence of peptides in the pharmaceutical business: From
  exploration to exploitation.
\newblock {\em {EuPA} Open Proteom.}, 4:58--69, September 2014.

\bibitem{Burroughes1990}
J.~H. Burroughes, D.~D.~C. Bradley, A.~R. Brown, R.~N. Marks, K.~Mackay, R.~H.
  Friend, P.~L. Burns, and A.~B. Holmes.
\newblock Light-emitting diodes based on conjugated polymers.
\newblock {\em Nature}, 347(6293):539--541, October 1990.

\bibitem{Koss2018}
Kyle Koss and Larry Unsworth.
\newblock Towards developing bioresponsive, self-assembled peptide materials:
  Dynamic morphology and fractal nature of nanostructured matrices.
\newblock {\em Materials}, 11(9):1539, August 2018.

\bibitem{Hartgerink2001}
J.~D. Hartgerink.
\newblock Self-assembly and mineralization of peptide-amphiphile nanofibers.
\newblock {\em Science}, 294(5547):1684--1688, November 2001.

\bibitem{Smith2008}
A.~M. Smith, R.~J. Williams, C.~Tang, P.~Coppo, R.~F. Collins, M.~L. Turner,
  A.~Saiani, and R.~V. Ulijn.
\newblock Fmoc-diphenylalanine self assembles to a hydrogel via a novel
  architecture based on $\pi$--$\pi$ interlocked $\beta$-sheets.
\newblock {\em Adv. Mater.}, 20(1):37--41, January 2008.

\bibitem{Grbitz2006}
Carl~Henrik G\"{o}rbitz.
\newblock The structure of nanotubes formed by diphenylalanine, the core
  recognition motif of alzheimer's $\beta$-amyloid polypeptide.
\newblock {\em Chem. Commun.}, 22:2332--2334, 2006.

\bibitem{frederix2015exploring}
Pim~WJM Frederix, Gary~G Scott, Yousef~M Abul-Haija, Daniela Kalafatovic,
  Charalampos~G Pappas, Nadeem Javid, Neil~T Hunt, Rein~V Ulijn, and Tell
  Tuttle.
\newblock Exploring the sequence space for (tri-) peptide self-assembly to
  design and discover new hydrogels.
\newblock {\em Nat. Chem.}, 7(1):30, 2015.

\bibitem{Chiti2003}
Fabrizio Chiti, Massimo Stefani, Niccol{\`{o}} Taddei, Giampietro Ramponi, and
  Christopher~M. Dobson.
\newblock Rationalization of the effects of mutations on peptide and protein
  aggregation rates.
\newblock {\em Nature}, 424(6950):805--808, August 2003.

\bibitem{Marchesan2012}
Silvia Marchesan, Christopher~D. Easton, Firdawosia Kushkaki, Lynne Waddington,
  and Patrick~G. Hartley.
\newblock Tripeptide self-assembled hydrogels: unexpected twists of chirality.
\newblock {\em Chem. Commun.}, 48(16):2195--2197, 2012.

\bibitem{Yap2010}
Chun~Wei Yap.
\newblock {PaDEL}-descriptor: An open source software to calculate molecular
  descriptors and fingerprints.
\newblock {\em J. Comp. Chem.}, 32(7):1466--1474, December 2010.

\bibitem{Wall2012}
Brian~D. Wall and John~D. Tovar.
\newblock Synthesis and characterization of $\pi$-conjugated peptide-based
  supramolecular materials.
\newblock {\em Pure Appl. Chem.}, 84(4):1039--1045, March 2012.

\bibitem{diamond1974interpretation}
Jared~M Diamond and Yehuda Katz.
\newblock Interpretation of nonelectrolyte partition coefficients between
  dimyristoyl lecithin and water.
\newblock {\em J. Membrane Biol.}, 17(1):121--154, 1974.

\bibitem{marrink1994simulation}
Siewert-Jan Marrink and Herman~JC Berendsen.
\newblock Simulation of water transport through a lipid membrane.
\newblock {\em J. Phys. Chem.}, 98(15):4155--4168, 1994.

\bibitem{orsi2010passive}
Mario Orsi and Jonathan~W. Essex.
\newblock Chapter 4 passive permeation across lipid bilayers: a literature
  review.
\newblock In {\em Molecular Simulations and Biomembranes: From Biophysics to
  Function}, pages 76--90. The Royal Society of Chemistry, 2010.

\bibitem{carpenter2014method}
Timothy~S Carpenter, Daniel~A Kirshner, Edmond~Y Lau, Sergio~E Wong, Jerome~P
  Nilmeier, and Felice~C Lightstone.
\newblock A method to predict blood-brain barrier permeability of drug-like
  compounds using molecular dynamics simulations.
\newblock {\em Biophys. J.}, 107(3):630--641, 2014.

\bibitem{lee2016simulation}
Christopher~T Lee, Jeffrey Comer, Conner Herndon, Nelson Leung, Anna Pavlova,
  Robert~V Swift, Chris Tung, Christopher~N Rowley, Rommie~E Amaro, Christophe
  Chipot, Yi~Wang, and James~C Gumbard.
\newblock Simulation-based approaches for determining membrane permeability of
  small compounds.
\newblock {\em J. Chem. Inf. Model.}, 56(4):721--733, 2016.

\bibitem{bennion2017predicting}
Brian~J Bennion, Nicholas~A Be, Margaret~Windy McNerney, Victoria Lao, Emma~M
  Carlson, Carlos~A Valdez, Michael~A Malfatti, Heather~A Enright, Tuan~H
  Nguyen, Felice~C Lightsto~ne, and Timothy~S Carpenter.
\newblock Predicting a drug's membrane permeability: A computational model
  validated with in vitro permeabi lity assay data.
\newblock {\em J. Phys. Chem. B}, 121(20):5228--5237, 2017.

\bibitem{tse2018link}
Chi~Hang Tse, Jeffrey Comer, Yi~Wang, and Christophe Chipot.
\newblock The link between membrane composition and permeability to drugs.
\newblock {\em J. Chem. Theory Comput.}, 14:2895--2909, 2018.

\bibitem{menichetti2017insilico}
Roberto Menichetti, Kiran~H Kanekal, Kurt Kremer, and Tristan Bereau.
\newblock $\textit {In silico}$ screening of drug-membrane thermodynamics
  reveals linear relations between bulk partitioning and the potential of mean
  force.
\newblock {\em J. Chem. Phys.}, 147(12):125101, 2017.

\bibitem{Rudzinski2019}
Joseph~F. Rudzinski.
\newblock Recent progress towards chemically-specific coarse-grained simulation
  models with consistent dynamical properties.
\newblock {\em Computation}, 7(3):42, August 2019.

\bibitem{menichetti2019revisiting}
Roberto Menichetti and Tristan Bereau.
\newblock Revisiting the {Meyer-Overton} rule for drug-membrane permeabilities.
\newblock {\em Mol. Phys.}, 117(20):2900--2909, 2019.

\bibitem{Centi2020}
Alessia Centi, Arghya Dutta, Sapun~H. Parekh, and Tristan Bereau.
\newblock Inserting small molecules across membrane mixtures: Insight from the
  potential of mean force.
\newblock {\em Biophys. J.}, 118(6):1321--1332, March 2020.

\bibitem{Cornell2017}
Caitlin~E. Cornell, Nicola~L.C. McCarthy, Kandice~R. Levental, Ilya Levental,
  Nicholas~J. Brooks, and Sarah~L. Keller.
\newblock n-alcohol length governs shift in {Lo-Ld} mixing temperatures in
  synthetic and cell-derived membranes.
\newblock {\em Biophys. J.}, 113(6):1200--1211, September 2017.

\end{thebibliography}

\end{document}